# Photo-enhanced antinodal conductivity in the pseudogap state of high-$T_c$ cuprates

F. Cilento[1*], S. Dal Conte[2,3,†], G. Coslovich[4,‡], S. Peli[2,5], N. Nembrini[2,5], S. Mor[2], F. Banfi[2,3], G. Ferrini[2,3], H. Eisaki[6], M.K. Chan[7], C. Dorow[7], M. Veit[7], M. Greven[7], D. van der Marel[8], R. Comin[9], A. Damascelli[9,10], L. Rettig[11,§], U. Bovensiepen[11], M. Capone[12], C. Giannetti[2,3*], F. Parmigiani[1,4].

[1] Elettra – Sincrotrone Trieste S.C.p.A., Basovizza I-34149, Italy.
[2] Department of Physics, Università Cattolica del Sacro Cuore, Brescia I-25121, Italy.
[3] i-Lamp (Interdisciplinary Laboratories for Advanced Materials Physics (i-LAMP), Università Cattolica del Sacro Cuore, Brescia I-25121, Italy.
[4] Department of Physics, Università degli Studi di Trieste, Trieste I-34127, Italy.
[5] Department of Physics, Università degli Studi di Milano, Italy.
[6] Nanoelectronics Research Institute, National Institute of Advanced Industrial Science and Technology, Tsukuba, Ibaraki 305-8568, Japan.
[7] School of Physics and Astronomy, University of Minnesota, Minneapolis, Minnesota 55455, USA.
[8] Département de Physique de la Matière Condensée, Université de Genève, Switzerland.
[9] Department of Physics and Astronomy, University of British Columbia, Vancouver, BC V6T 1Z1, Canada.
[10] Quantum Matter Institute, University of British Columbia, Vancouver, BC V6T 1Z4, Canada.
[11] Fakultaet fuer Physik and Zentrum für Nanointegration (CENIDE), Universitaet Duisburg-Essen, 47048 Duisburg, Germany.
[12] CNR-IOM Democritos National Simulation Center and Scuola Internazionale Superiore di Studi Avanzati (SISSA), Via Bonomea 265, 34136 Trieste, Italy.

* Correspondence and requests for materials should be addressed to F.C. (email: federico.cilento@elettra.eu) or to C.G. (email: claudio.giannetti@unicatt.it).
† Present address: Department of Physics, Politecnico di Milano, I-20100 Milano, Italy.
‡ Present address: Materials Sciences Division, Lawrence Berkeley National Laboratory, Berkeley, CA 94720, USA.
§ Present address: Swiss Light Source, Paul Scherrer Institute, Villigen, Switzerland

**A major challenge in understanding the cuprate superconductors is to clarify the nature of the fundamental electronic correlations that lead to the pseudogap phenomenon. Here we use ultrashort light pulses to prepare a non-thermal distribution of excitations and capture novel properties that are hidden at equilibrium. Using a broadband (0.5-2 eV) probe we are able to track the dynamics of the dielectric function, unveiling an anomalous decrease of the scattering rate of the charge carriers in a pseudogap-like region of the temperature ($T$) and hole-doping ($p$) phase diagram. In this region, delimited by a well-defined $T^*_{neq}(p)$ line, the photo-excitation process triggers the evolution of antinodal excitations from gapped (localized) to delocalized quasi-particles characterized by a longer lifetime. The novel concept of photo-enhanced antinodal conductivity is naturally explained within the single-band Hubbard model, in which the short-range Coulomb repulsion leads to a k-space differentiation between "nodal" quasiparticles and antinodal excitations.**

Superconductivity in copper oxides takes place when charge carriers are injected into a charge-transfer insulator (1), in which the carriers are localized by the strong electron-electron interactions. This surprising phenomenon has motivated an incredible effort to understand to what extent the electron-electron interactions determine the physical properties of cuprates when tuning these materials away from the insulating state by increasing the hole concentration $p$ and the temperature $T$. The physics of superconducting cuprates is further complicated by the ubiquitous opening, even above the superconducting critical temperature, of an anisotropic gap (pseudogap) in the electronic density of states, having the maximal value close to the $\mathbf{k}=(\pm\pi,0), (0,\pm\pi)$ regions (antinodes) of the Brillouin zone (2). In the same region of the $p$-$T$ phase diagram, different ordered phases have been observed. Such phases can be favored by the reduced kinetic energy of the carriers and might be associated with a quantum critical point underneath the superconducting dome (3, 4). Indeed, a wealth of different broken-symmetries, such as unusual q=0 magnetism (5–7), charge density waves (8, 9), stripes (10), nematic and smectic phases (11-13), have been reported. The universal physical mechanism underlying the pseudogap of cuprates continues to be subject of intense research (14) and the relation between the electronic interactions and the various ordering tendencies remains one of the major open questions.

A possible scenario to reconcile this phenomenology is that the pseudogap emerges as an inherent effect of the strong short-range Coulomb repulsion, $U$, between two electrons occupying the same lattice site. Considering the case of an isotropic Mott insulator, the $U$-driven suppression of charge fluctuations is expected to reduce the electron kinetic energy and render the electronic excitations more and more localized in real space. Cluster generalization of Dynamical Mean Field Theory (DMFT) calculations (15–17) suggests that copper oxides exhibit a similar $U$-driven reduction of kinetic energy that is, however, not uniform in momentum space. In the underdoped region, the antinodal excitations are indeed quasi-localized, with a $T$=0 divergent scattering rate that is reminiscent of the Mott insulator (see Fig. 1a), whose self-energy diverges at low frequency and it is reduced at finite temperature. In contrast, the "nodal" excitations in the vicinity of $\mathbf{k}=(\pm\pi/2,\pm\pi/2)$ still retain the essential nature of the quasiparticles (QPs) at large hole concentrations, i.e., a scattering rate that increases when external energy is provided (see Fig. 1b). The possible $\mathbf{k}$-space differentiation between nodal QPs and antinodal excitations in the pseudogap phase has been hitherto elusive both to $\mathbf{k}$-space integrated equilibrium techniques, such as optical spectroscopy and resistivity measurements, that are mostly sensitive to the properties of nodal QPs, and to conventional Angle Resolved Photoemission Spectroscopy (ARPES), that has limits in capturing small temperature- and $\mathbf{k}$-dependent variations of the electronic scattering rate.

Here, we adopt a non-equilibrium approach (18) based on ultrashort light pulses (≈100 fs) used to "artificially" prepare the system in a non-thermal state with an excess of antinodal (AN) excitations (19). The key element of our experiment, which goes beyond single-color pump-probe techniques [20-22], is that the dynamics of the optical properties are simultaneously probed over an unprecedentedly broad energy range (0.5-2 eV). This technique allows us to probe the damping of the infrared reflectivity plasma edge and directly relate the transient reflectivity variation, $\delta R(\omega,t)$, to the instantaneous value of the optical scattering rate. From the outcome of time-resolved broadband spectroscopy, we infer that the transient non-thermal state created in the pseudogap phase is characterized by a scattering rate smaller than that at equilibrium. The picture emerging is that, upon excitation, the localized antinodal states transiently evolve into more mobile states with a reduced scattering rate. This scenario is corroborated by Dynamical Mean Field Theory calculations within the single-band Hubbard model and by time-resolved ARPES data in the pseudogap state. Finally, the generality of the concept of photo-enhanced antinodal conductivity in the pseudogap is demonstrated by combining the results obtained on different families of copper oxides (Bi- and Hg-based) in a single and universal phase diagram.

**RESULTS:**

Non-equilibrium optical spectroscopy is emerging as a very effective tool to unravel the different degrees of freedom coupled to electrons in correlated materials (18,23). After the photoexcitation process, small variations of the equilibrium optical properties can be measured on a timescale faster than the recovery of the equilibrium QPs distribution, the complete restoration of the (long-range) orders (24) and the heating of the phonons (25), thus unveiling an intriguing physics that cannot be accessed in equilibrium conditions. Single-color time-resolved reflectivity measurements (20-22) have been applied in the past to study the pseudogap state, evidencing a characteristic dynamics proportional to the pseudogap amplitude ($|\Delta_{pg}|$) and a change of sign of the photoinduced reflectivity variation at $T\approx T^*$. Nonetheless, the inherent lack of spectral information of single-color techniques precluded the understanding of the origin of this pseudogap-related reflectivity signal. Recent developments in ultrafast techniques now permit to overcome this limitation by probing the dynamics of the dielectric function over a broad frequency range. This paves the way to a quantitative modeling of the ultrafast optical response of the pseudogap phase.

**Optical properties of doped-cuprates and the Extended Drude Model**

To properly model the dynamics of the optical properties measured in non-equilibrium conditions, we start introducing the basic elements that characterize the equilibrium reflectivity, $R_{eq}(\omega)$, of optimally doped $Bi_2Sr_2Y_{0.08}Ca_{0.92}Cu_2O_{8+\delta}$ (OP-YBi2212, hole doping $p=0.16$, $T_c=96$ K), shown in Figure 2a. In the infrared region, the normal-incidence $R_{eq}(\omega)$ is dominated by a metallic-like response, which is characterized by a broad edge at the dressed plasma frequency $\omega_p\approx 1.25$ eV. The infrared reflectivity of optimally and overdoped cuprates is well reproduced by the Extended Drude Model (EDM) (26,27) in which a frequency-dependent optical scattering time, $\tau(\omega,T)$, accounts for all the processes that affect the QP lifetime, such as the electronic scattering with phonons, spin fluctuations or other bosons of electronic origin. Within the EDM the optical scattering rate is connected to the single-particle self-energy (SE) by:

$$\gamma(\omega,T) = \frac{\hbar}{\tau(\omega,T)} = \text{Im}\left\{\omega\left[\int_{-\infty}^{+\infty}\frac{f(\zeta,T)-f(\zeta+\omega,T)}{\omega+\Sigma^*(\zeta,T)-\Sigma(\zeta+\omega,T)}d\zeta\right]^{-1}-\omega\right\} \quad (1)$$

where $f$ is the Fermi-Dirac distribution and $\Sigma(\zeta,T)$ and $\Sigma^*(\zeta,T)$ are the electron and hole k-space averaged SEs. In Figure 2a we report the best fit of the EDM to the experimental $R_{eq}(\omega)$ of OP-YBi2212 at $T$=100 K. In the calculation of the SE in Eq. 1, we use a recently developed model (28-30), that takes into account a non-constant QPs density of states, $\widetilde{N}(\omega,T)$, characterized by a pseudogap width $\Delta_{pg}$=40 meV (see Methods section). The k-space integrated electronic Density of States (DOS) is recovered completely between $\Delta_{pg}$ and $2\Delta_{pg}$. The $\widetilde{N}(\omega,T)$ extracted by optical spectroscopy (see Methods, Supplementary Notes 1,2, and Supplementary Figures 1,2) is in agreement with the outcome of tunneling experiments (31). The quality of the fit to the optical properties (Fig. 2a) demonstrates that the EDM with a non-constant DOS is a very effective tool to extract the scattering properties of the charge carriers, at least for moderate hole-doping concentrations. The possible failure of the EDM model is expected for hole-doping concentration smaller than those of the samples studied in this work. The interband transitions at $\hbar\omega$>1.5 eV are accounted for by additional Lorentz oscillators at visible/UV frequencies. Focusing on the energy range that will be probed by the time-resolved experiment (0.5-2 eV), we note that $\tau(\omega)$ has almost approached the asymptotic value $\tau_\infty$≈2 fs (see inset of Fig. 2a). Therefore, we can safely assume that, in the 0.5-2 eV range, the damping of the reflectivity edge depends on the asymptotic value of the scattering rate, $\gamma_\infty=\hbar/\tau_\infty$.

**Non-equilibrium optical spectroscopy**

In the non-equilibrium optical spectroscopy, snapshots of the reflectivity edge of OP-YBi2212 are taken with 100 fs time resolution, as a function of the delay from the excitation with the 1.5 eV pump pulses. Considering that the temporal width of the probe pulses is much longer than the inherent scattering time of the charge carriers ($\tau_\infty$≈2 fs), we can assume that the electronic excitations have completely lost any coherence on a timescale faster than the observation time. Therefore, the transient optical properties can be rationalized by changing some effective parameters in the equilibrium reflectivity. Notably, $R_{eq}(\omega, \gamma_\infty)$ exhibits an isosbestic point (see Fig. 2a) at the frequency $\widetilde{\omega} \sim 1.1$ eV. At this frequency, the reflectivity is independent of $\gamma_\infty$ and, for small $\delta\gamma_\infty$ changes, it can be expanded as $\delta R(\omega,\gamma_\infty)=[\partial R(\omega)/\partial\gamma_\infty]\delta\gamma_\infty$, where $[\partial R(\omega)/\partial\gamma_\infty]>(<)0$ for $\omega>(<)\widetilde{\omega}$ (see Supplementary Note 3 and Supplementary Fig. 3). Therefore, the reflectivity variation, measured over a broad frequency range across $\widetilde{\omega}$, can provide a direct information about the instantaneous value of the total scattering rate during the thermalization process of the photoinduced non-equilibrium QP population. In the bottom panel of Figure 2a we report the normalized reflectivity variation $\delta R/R(\omega)$, calculated from the equilibrium EDM model by assuming a positive (red line) or negative (blue line) variation of the total scattering rate, $\gamma_\infty$. The reflectivity variation associated to a change in $\gamma_\infty$ extends over a broad frequency range and it changes sign when crossing the isosbestic point $\widetilde{\omega}$.

Fig. 2b shows the frequency-resolved measurements on OP-YBi2212 at $T$=100 K, as a function of the delay t from the pump pulse. The relative reflectivity variation $\delta R/R(\omega,$ t$)=[R_{neq}(\omega,t)-R_{eq}(\omega)]/R_{eq}(\omega)$, where $R_{neq}$ is the non-equilibrium reflectivity, is reported as a two-dimensional plot. The colour scale represents the magnitude of $\delta R/R(\omega,t)$, as a function of t and of the probe photon energy $\hbar\omega$. Soon after the excitation, $\delta R/R(\omega,t)$ is positive below $\widetilde{\omega}$, whereas a negative signal is detected above $\widetilde{\omega}$. When compared to the $\delta R/R(\omega)$ calculated for $\delta\gamma_\infty<0$ (bottom panel of Figure 2a), the measured $\delta R/R(\omega,t)$ suggests a transient decrease of the QP scattering rate soon after the excitation with the pump pulse.

To substantiate the possible transient variation of $\gamma_\infty$, a differential model is used to quantitatively analyze the $\delta R/R(\omega,t)$ signal. This approach consists in finding out the minimal set of parameters that should be changed in the equilibrium dielectric function to satisfactorily reproduce the measured $\delta R/R(\omega,t)$ at a given time t. The main goal of this procedure is to disentangle the contributions of the genuine variation of $\gamma_\infty$, from that of possible photo-induced band-structure modifications, such as the transient filling of the pseudo-gapped electronic states (30,32). In Figure 3 we report a slice of $\delta R/R(\omega,t)$ (red dots) at fixed delay time, i.e. t=100 fs. $\delta R/R(\omega,t=100\text{ fs})$ is qualitatively reproduced, over the whole probed frequency range, by modifying only two parameters in the equilibrium EDM, i.e., $\widetilde{N}(\omega,T)$ and $\gamma_\infty$. In contrast, no change of the interband transitions and of the plasma frequency is required. This demonstrates that, in the probed spectral range, the $\delta R/R(\omega)$ signal is not significantly affected by the transient change of the electronic occupation at the pump energy scale or by other excitonic-like processes and that the density of the charge carriers is not significantly modified by the excitation process. Similarly, Fig. 3 also displays the contributions to the $\delta R/R(\omega,t=100\text{ fs})$ signal, as arising from the variation of $\widetilde{N}(\omega,T)$ (green line) and $\gamma_\infty$ (blue line), separately. The central result is that a negative $\gamma_\infty$ variation is necessary to reproduce $\delta R/R(\omega,t=100\text{ fs})$. $\delta\gamma_\infty<0$ corresponds to a narrowing of the reflectivity plasma edge around the isosbestic point at $\widetilde{\omega}$ (see Figure 2a), that is detected as a broad reflectivity variation changing from positive to negative when moving to high frequencies. Quantitatively, the relative variation $\delta\gamma_\infty/\gamma_\infty=-(1.2\pm0.4)\cdot10^{-2}$ is extracted from the fitting procedure. For sake of clarity, we show in Figure 3 the total $\delta R/R(\omega)$ calculated (black dashed line) by constraining a positive $\delta\gamma_\infty/\gamma_\infty=+1.2\cdot10^{-2}$ variation (red dashed line). In this case, the main features of the measured $\delta R/R(\omega)$ cannot be, even qualitatively, reproduced. This demonstrates that while the experimental uncertainties can affect the absolute value of $\delta\gamma_\infty/\gamma_\infty$, the transient photoinduced decrease of the scattering rate is a robust experimental fact that is unaffected by possible uncertainties, such as fluence fluctuations or changes in the pump probe overlap.

The characteristic relaxation time $\tilde{t}$ of the measured $\delta\gamma_\infty$ is obtained by fitting the function $\delta R/R(t)=\delta R/R(0)\exp(-t/\tilde{t})$ to the time-resolved traces at fixed frequencies. In particular, we focus on the time-traces at $\hbar\omega$ =0.68 eV and 1.55 eV, shown in Fig. 3b and 3c, for which the contribution from the $\tilde{N}(\omega,T)$ variation is negligible. The resulting value, $\tilde{t}$ =600±50 fs (see Supplementary Note 4 and Supplementary Fig. 4), is of the same order as the time required for the complete heating of the lattice and for the recovery of the equilibrium QPs population, as will be shown by the time-resolved photoemission measurements. Interestingly, on the picosecond timescale the $\delta R/R(\omega)$ signal is qualitatively opposite to the signal at t≈100 fs, as shown by the time-traces in Fig. 3b and 3c. In particular, it changes from negative to positive as $\omega$ increases and crosses $\tilde{\omega}$. The $\delta R/R(\omega,t>1$ ps) signal can be reproduced by a broadening of the reflectivity edge, equivalent to an increase of the scattering rate ($\delta\gamma_\infty/\gamma_\infty=10^{-4}$), with a negligible contribution from the variation of $\tilde{N}(\omega,T)$. This $\gamma_\infty$ increase is compatible with a local effective heating ($\delta T$=0.6 K) of the lattice, that can be estimated considering the pump fluence (10 μJ/cm$^2$) and the specific heat of the sample (see Supplementary Note 5 and Supplementary Fig. 5).

The transient decrease of the total electronic scattering rate, measured on the sub-ps timescale at $T$=100 K, is a novel finding that contrasts both with the results (25) obtained well above the pseudogap temperature and with the behavior expected for a metallic system. In the latter case, the scattering rate should increase as energy is delivered, either adiabatically or impulsively. Furthermore, as inferred from the $\delta R/R(\omega,t)$ measured in local quasi-equilibrium conditions (t>1 ps), this anomalous $\delta\gamma_\infty<0$ cannot originate from the partial quench of an order parameter or a temperature-induced decrease of the electron-boson coupling. In fact, the expected positive $\delta\gamma_\infty>0$ is recovered at t>1 ps, although on this timescale an increase of the effective electronic and bosonic temperatures and the related amplitude decrease of the possible order parameter is attained.

Non-equilibrium broadband optical spectroscopy crucially broadens the information that could be reached in the past by single-color techniques. The quantitative modeling of the dynamics of the dielectric function in the 0.5-2 eV energy range demonstrates that, after the impulsive photo-excitation, the optimally-doped YBi2212 sample at 100 K is driven into a non-equilibrium state, characterized by a scattering rate smaller than that at equilibrium ($\delta\gamma_\infty$<0). In terms of the AC conductivity, this corresponds to a transient increase of the conductivity, once assumed a constant density of the charge carriers. The relaxation time of this photo-enhanced conductivity is $\tilde{t}$ =600±50 fs. On the picosecond timescale the equilibrium distribution, dominated by nodal QPs, is recovered and a more conventional behavior of the scattering rate ($\delta\gamma_\infty$>0) is observed, in agreement with the outcomes of conventional equilibrium techniques, such as resistivity and optics (33,34).

**The k-space dependent non-equilibrium QP population**

Since the energy scale of the pump photons (1.5 eV) is much larger than the thermal ($k_BT$≈10 meV) and the pseudogap ($\Delta_{pg}$≈40 meV) energy scales, the photo-induced QPs distribution is expected to have no relation with the equilibrium one. The photoexcitation process can be roughly reduced to two main steps. In the first step, the pump pulse is absorbed creating electron-hole excitations extending from -1.5 eV to +1.5 eV across the Fermi energy ($E_F$) and with a distribution that is regulated by the Joint Density of States of the photo-excitation process. In consequence of the extremely short scattering time of high-energy excitations ($\tau_\infty=\hbar/\gamma_\infty$≈2 fs), this photoexcited population undergoes a fast energy relaxation related to multiple scattering processes that lead to the creation of excitations at energies closer to $E_F$. Considering the **k**-space integrated density of states in the pseudogap state, a high density of excitations is expected to accumulate, within the pump pulse duration (0-100 fs), in the $\Delta_{pg}$-$2\Delta_{pg}$ energy range from $E_F$. In the second and slower step, the subset of scattering processes that allow large momentum exchange, while preserving the energy conservation, leads to the recovery of a quasi-equilibrium distribution dominated by nodal QPs. Considering the phase-space constraints for these processes, the redistribution of excitations in the **k**-space is expected to be effective on the picosecond timescale.

To directly investigate the **k**-space electron distribution after the photoexcitation process with 1.5 eV ultrashort pulses, we apply a momentum-resolved technique. Time- and angle-resolved photoemission spectroscopy (TR-ARPES, see Methods section) is applied to measure the transient occupation at different **k**-vectors (35,36) along the Fermi arcs (see Fig. 4a) of a $Bi_2Sr_2CaCu_2O_{8+\delta}$ single crystal at $T$=100 K. When considering the TR-ARPES spectra at fixed angles $\Phi$ from the antinodal direction (see Fig. 4a), the pump excitation results in a depletion of the filled states below $E_F$ and a filling of the states above $E_F$ (35). In the following, we will focus on the integral ($I$) of the TR-ARPES spectra for $E>E_F$, that is proportional to the total excess of excitations in the empty states. Fig. 4b reports the variation of $I$ normalized to the intensity before the arrival of the pump pulse ($\Delta I_\Phi(t)/I_\Phi$), as a function of $\Phi$. For each value of $\Phi$, $\Delta I_\Phi(t)/I_\Phi$ is fitted by a single exponential decay of the form: $\Delta I_{\Phi 0}\exp(-t/\tilde{t}_\Phi)$, convoluted with a Gaussian function to account for the finite temporal resolution (~100 fs). Fig. 4c and 4d show the **k**-dependent values of $\Delta I_{\Phi 0}$ and $\tilde{t}_\Phi$ extracted from the fitting procedure. Within the error bars, $\Delta I_{\Phi 0}$ increases of a factor 3 when moving from the node ($\Phi$=45°) towards the antinodal region of the BZ, indicating an effective photoinjection of AN excitations. This situation is dramatically different from that expected in equilibrium conditions, in which the number of excitations is governed solely by the Fermi-Dirac distribution at temperature $T$. In this case, the excitations density should significantly decrease when approaching the AN region of the Brillouin zone, as a consequence of the gap $\Delta_{pg} \gg k_B T$ in the density of electronic states. Another notable result is that the relaxation time $\tilde{t}_\Phi$ is **k**-dependent, increasing from 300 fs to 800 fs when moving from the N to the AN region. These values are much larger than the total optical scattering time ($\tau_\infty = \hbar/\gamma_\infty \approx 2$ fs), which demonstrates that the recovery of a quasi-equilibrium QPs distribution is severely constrained, either by the phase space available for the scattering processes simultaneously conserving energy and momentum, or by some bottleneck effect related to the emission of gap-energy bosons during the relaxation of AN excitations.

TR-ARPES demonstrates that, soon after the excitation with 1.5 eV pump pulses, an excess number of low-energy electron excitations is accumulated in the antinodal region. Furthermore, the relaxation time of this non-equilibrium distribution is significantly **k**-dependent. Taken together, these results prove that this non-equilibrium electron distribution in the **k**-space cannot be described by a Fermi-Dirac distribution with a single effective electronic temperature that is evolving in time. At the simplest level, we can assume that the "effective" temperature increase of the excitations in the antinodal region is larger than that of the nodal QP population. Notably, the **k**-space averaged relaxation time of the non-equilibrium distribution, $\langle \tilde{t}_\Phi \rangle = 550 \pm 200$ fs, is the same, within the error bars, of the relaxation time of the transient $\delta\gamma_\infty < 0$ measured by non-equilibrium optical spectroscopy. This result demonstrates the direct relation between the creation of a non-thermal distribution with an excess of antinodal excitations and the transient decrease of the scattering rate measured by non-equilibrium optical spectroscopy.

## The $T^*_{neq}(p)$ line emerging from non-equilibrium spectroscopy

The use of ultrashort light pulses to manipulate the equilibrium QP distribution is crucial to investigate the cuprate phase diagram from a perspective that was hitherto inaccessible. In this section we report on the $T^*_{neq}(p)$ temperature below which the transient reduction of $\gamma_\infty$ is observed via non-equilibrium optical spectroscopy. As shown in Figure 3a, for $\omega$ significantly above $\omega_p$ the contribution to $\delta R/R(\omega, t)$ due to the modification of $\tilde{N}(\omega)$ is negligible. Therefore, single-colour pump-probe measurements at 1.55 eV probe photon energy contain the direct signature of the transient $\delta\gamma_\infty<0$, in the form of a negative component (20,21) with relaxation time $\tilde{t}$ =600 fs (22). Fig. 5a reports some of the single-colour time-traces measured as a function of the temperature (from 300 K to 20 K) for different hole concentrations. The negative component (blue colour) appears in the $\delta R/R(t)$ signal at $T^*_{neq}$=240±20 K for $p$=0.13 and at $T^*_{neq}$=165±20 K for $p$=0.16. Above $p$=0.18, this negative $\delta R/R(t)$ signal is never detected upon cooling the sample down to $T_c$ (see Supplementary Notes 6,7 and Supplementary Figures 6,7,8). A similar behavior is also found for underdoped $HgBa_2CuO_{4+\delta}$ (Hg1201) (see Supplementary Note 8 and Supplementary Figures 9,10,11). Hg1201 is considered a nearly ideal single-layer cuprate and exhibits a maximal critical temperature close to that of double-layer YBi2212 (37,38). Collecting the results for the two materials on the same plot, we obtain a phase diagram that shows a universal behavior for the properties of different copper-oxide based superconductors having the same maximal critical temperature. Fig. 5b shows that the $p$-$T$ phase diagram of cuprates is dominated by an ubiquitous and sharp $T^*_{neq}(p)$ boundary. Below this line, the pump-induced non-thermal distribution of the charge carriers exhibits a scattering rate smaller than that at equilibrium. This pseudogap-like $T^*_{neq}(p)$ boundary meets the superconducting dome slightly above $p$=0.18. Below $T_c$, the continuation of the $T^*_{neq}(p)$ line delimits two regions of the superconducting dome that exhibit opposite variation of the optical spectral weight of

intra- and inter-band transitions (23,39-41), related to the crossing from a kinetic energy gain- to a potential energy gain-driven superconducting transition, consistent with predictions for the two-dimensional Hubbard model (42).

Further insight into the $T^*_{neq}(p)$-line unveiled by non-equilibrium optical spectroscopy is provided by the comparison with the pseudogap temperature $T^*(p)$ estimated from complementary equilibrium techniques (7,43,44,14) (elastic neutron scattering, resistivity, resonant ultrasound spectroscopy) and with the onset temperature of different ordered states (45-47) (charge-density wave, time-reversal symmetry breaking states and fluctuating stripes). The picture sketched in Fig. 5c compares the main phenomenology on the most common materials with similar critical temperatures (Bi2212, Hg1201, YBCO). Remarkably, the $T^*_{neq}(p)$ line almost exactly coincides with the pseudogap line estimated by resistivity measurements (44) and ultrasound spectroscopy (14) and with the onset of the newly-discovered q=0 exotic magnetic modes (7,43). The appearance of ordered states at $T<T^*$ is likely the consequence of the instability of the correlated pseudogap ground state upon further cooling.

**The antinodal scattering rate as the signature of short-range Coulomb repulsion**

The results reported in the previous sections suggest that the $T^*_{neq}(p)$ line delimits a region in which the AN states evolve into more metallic ones ($\delta\gamma_\infty<0$) upon photo-excitation with the pump pulses. The generality of the results obtained calls for a general model that accounts for the phase diagram unveiled by the non-equilibrium optical spectroscopy. Considering that the measured transient decrease of the carrier scattering rate is faster than the complete heating of the lattice, we focus on the minimal model that neglects electron-phonon coupling and retains the genuine physics of correlations, i.e., the two-dimensional

Hubbard Hamiltonian (48,49). In order to compute the temperature dependent self-energy in different positions of the Brillouin zone we use the Dynamical Cluster Approximation (DCA), a Cluster extension of Dynamical Mean-Field Theory that captures the **k**-space differentiation of the electronic properties (16) between different regions of the Brillouin zone (see Methods section). Furthermore, long-range correlations are neglected in order to focus on the intrinsic effect of short-range correlations inside the chosen 4-site cluster (see Methods section). At this stage, the increase of energy related to the pump excitation is mimicked by selectively increasing the "effective" temperature of the nodal and antinodal self-energies.

The FS of a lightly doped system is reported in Fig. 6a and exhibits a progressive smearing when moving from the nodes to the antinodes. This result is qualitatively in agreement with the FS experimentally measured by conventional ARPES in prototypical cuprates (compare with the ARPES data on optimally-doped $Bi_2Sr_2Y_{0.08}Ca_{0.92}Cu_2O_{8+\delta}$ at 100 K reported in Fig. 4a). The smearing of the FS at antinodes reflects a strong dichotomy between the scattering rate of nodal and antinodal excitations that can be captured by plotting the imaginary parts of the calculated SE, i.e. the inverse QP lifetime, as a function of the effective temperature (see Fig. 6c). In contrast to nodal QPs, whose scattering rate increases with temperature, the scattering rate of antinodal excitations exhibits a completely different evolution, decreasing as the effective temperature rises. This anomalous behavior is related to the localized and gapped character of the antinodal fundamental excitations that experience very strong electronic interactions at low temperatures. The concept of delocalized QPs, characterized by a smaller scattering rate, is progressively recovered when the temperature increases. This striking dichotomy of the nature of the elementary excitations in the **k**-space is the consequence of a momentum-space selective opening of a correlation-driven gap, which eventually evolves into the full Mott gap at $p=0$. The same physics has been previously identified in calculations based on similar approaches (also using larger clusters to achieve a better momentum resolution) (15–17, 50). When the hole doping is increased (see Fig. 6b and 6d) the **k**-space differentiation of N-AN fundamental excitations is washed out and a more conventional metallic behaviour is recovered. In this case, the delocalized QPs exhibit a gapless energy spectrum and a scattering rate that is proportional to a power function of the temperature over the entire Brillouin zone.

DMFT calculations thus confirm an intrinsic $U$-driven momentum-space differentiation of the electronic properties of cuprates at finite hole concentrations and temperatures. The nature of the AN states is similar to that of a Mott insulator in the sense that the scattering rate of AN states decreases when the internal energy of the system is increased.

**DISCUSSION:**

Wrapping up the experimental outcomes collected in this work, we can gain a novel insight into the pseudogap physics of high-$T_c$ cuprates. Time-resolved optical spectroscopy demonstrates that, upon excitation with 1.5 eV pump pulses, the optical properties of cuprates transiently evolve into those of a more conductive system ($\delta\gamma_\infty<0$). TR-ARPES shows that the **k**-space distribution of the electron excitations created by the pump pulse is characterized by an excess of AN electron. Finally, the reported observation of a transient enhancement of the conductivity unveils a universal and sharp pseudogap-like $T^*_{neq}(p)$ boundary in the $p$-$T$ phase diagram. All these results cannot be rationalized in terms of the simple consequence of an anisotropic scattering of QPs with bosonic fluctuations, such as antiferromagnetic (AF) spin fluctuations. Although AF correlations may strongly influence the dynamics of AN excitations, the energy provided by the pump pulse should necessary result in an increase of the boson density, leading to an increase of the scattering rate for all the timescales. Furthermore, the transient decrease of the scattering rate on the sub-ps timescale cannot be related to the presence of incipient charge orders, such as charge-density waves (CDW), that are quenched by the pump pulse. The picosecond dynamics of conventional CDW has been widely studied in weakly-correlated ($U\approx0$) materials (24,51). After the impulsive excitation, the characteristic timescale for the CDW recovery is on the order of several picosecond, i.e., much longer than the transient $\delta\gamma_\infty<0$ measured in our work. Even assuming a purely electronic (and faster) density wave mechanism in cuprates, the photoinduced decrease of the scattering rate should monotonically decrease eventually approaching a zero value. This is in contrast with the experimental observation of a transition from $\delta\gamma_\infty<0$ to $\delta\gamma_\infty>0$ on the picosecond timescale.

On the other hand, the Hubbard model provides a minimal framework for the interpretation of the pseudogap region of cuprates as unveiled by time-resolved spectroscopies. The pump pulse provides energy to the system in a non-thermal way, which can be schematized as a larger increase of the effective temperature of AN excitations as compared to that of nodal QPs. The transient decrease of the scattering rate ($\delta\gamma_\infty<0$) is thus related to the evolution of AN excitations from Mott-like gapped excitations to delocalized QPs with a longer lifetime. On the picosecond timescale the equilibrium electronic distribution is recovered and the expected $\delta\gamma_\infty>0$ is measured. Although this picture does not exhaust all the properties of the pseudogap, it captures a key element of the universal and fundamental nature of the antinodal states, providing a backbone for more realistic multi-band descriptions (3) that could give rise to a broken-symmetry state originating from a quantum phase transition (4) at $T=0$. Furthermore, the short-range Coulomb repulsion induces a suppression of the charge fluctuations below the $T^*_{neq}(p)$ line that is opposite to the effect of temperature. Therefore, the region of the cuprate phase diagram delimited by $T^*_{neq}(p)$ is intrinsically prone to bulk (52,53) and surface (54,55) phase-separated instabilities, whose nature depends on the details of the Fermi surface of the particular system considered. Interestingly, the region of the cuprate phase diagram in which the optical photoexcitation creates a non-equilibrium state with longer lifetime, closely corresponds to that in which the possibility of creating a transient superconductive state by THz excitation has been recently discussed (56). These findings suggest a general tendency of copper oxides to develop, when photoexcited, a transient non-equilibrium state that is more conductive than the equilibrium phase.

**Methods**

**Samples**
The Y-substituted Bi2212 single crystals were grown in an image furnace by the travelling-solvent floating-zone technique with a non-zero Y content in order to maximize $T_c$ (37). The underdoped samples were annealed at 550 °C for 12 days in a vacuum-sealed glass ampoule with copper metal inside. The overdoped samples were annealed in a quartz test tube under pure oxygen flow at 500 °C for 7 days. To avoid damage of the surfaces, the crystals were embedded in $Bi_2Sr_2CaCu_2O_{8+\delta}$ powder during the annealing procedure. In both cases, the quartz tube was quenched to ice-water bath after annealing to preserve the oxygen content at annealing temperature.

The Hg1201 single crystals were grown using a flux method, characterized, and heat treated to the desired doping level (38). The crystal surface is oriented along the *ab* plane with a dimension of about 1 mm$^2$. Hg1201 samples are hygroscopic. Therefore, the last stage of the preparation of the sample surface is done under a continuous flow of nitrogen, upon which the sample is transferred to the high-vacuum chamber ($10^{-7}$ mbar) of the cryostat within a few minutes. Before each measurement the surface is carefully checked for any evidence of oxidation.

**Optical Spectroscopy**
The *ab*-plane dielectric function at equilibrium of the Y-Bi2212 samples has been measured by conventional spectroscopic ellipsometry (57). The dielectric function has been obtained by applying the Kramers-Kronig relations to the reflectivity for $50<\omega/2\pi c<6000$ cm$^{-1}$ and directly from ellipsometry for $1500<\omega/2\pi c<36000$ cm$^{-1}$.

The $\delta R/R(\omega, t)$ data presented in Fig. 2 have been acquired combining two complementary techniques: i) A pump supercontinuum-probe setup (58), based on the white light generated in a Photonic Crystal Fiber seeded by a Ti:Sapphire cavity-dumped oscillator, to explore the visible-near infrared range of the spectrum (1.1-2 eV); ii) A pump tunable-probe setup, based on an Optical Parametric Amplifier (OPA) seeded by a Regenerative Amplifier, to extend measurements in the infrared spectral region (0.5-1.1 eV). In both cases, the laser systems operate at 250 kHz repetition rate. The pump fluence is set to 10±2 µJ/cm$^2$ for spectroscopic measurements. Single-color measurements ($\hbar\omega$ =1.55 eV), presented in Fig. 5, have been performed directly using the output of a cavity-dumped Ti:Sapphire oscillator. The high frequency modulation of the pump beam, combined with a fast scan of the pump-probe delay and lock-in acquisition, ensure a high signal-to-noise ratio (~10$^6$) and fast acquisition times, necessary to study the evolution of the time-resolved optical properties as a function of the temperature. Single-color measurements have been performed with a pump fluence ranging from 3 to 30 µJ/cm$^2$. In all experiments, the pump photon energy is 1.55 eV. Samples are mounted on the cold finger of a closed-cycle cryostat. The temperature of the sample is stabilized within ±0.5 K.

**The Extended Drude Model with a Non-Constant Density-of-States**

In the Extended Drude Model (EDM) the scattering processes are effectively accounted for by a temperature- and frequency-dependent scattering rate $\gamma(\omega,T)$, which is often expressed through the so-called "Memory Function", $M(\omega,T)$. The dielectric function resulting from the EDM is:

$$\varepsilon_D(\omega,T) = 1 - \frac{\omega_p^2}{\omega(\omega + M(\omega,T))}$$

In the conventional formulation of the EDM (for more details, see (27,30)), the calculation of the self-energy $\Sigma(\omega,T)$ is based on the assumption of a constant density of states at the Fermi level. This approximation is valid at $T$=300 K in optimally and over-doped systems, but fails as the temperature and the doping decrease and a pseudogap opens in the electronic density of states. A further evolution of the EDM, accounting for a non-constant electronic density of states, has been recently developed (28), and has been used to analyze spectroscopic data at equilibrium (29). Within this model, the imaginary part of the electronic self-energy is given by:

$$\Sigma_2(\omega,T) = -\pi \int_0^\infty \Pi(\Omega)\left\{\widetilde{N}(\omega+\Omega,T)[n(\Omega,T)+f(\omega+\Omega,T)] + \widetilde{N}(\omega-\Omega,T)[1+n(\Omega,T)-f(\omega-\Omega,T)]\right\}d\Omega$$

where $n, f$ are the Bose-Einstein and Fermi-Dirac distribution functions, respectively; $\Pi(\Omega)$ is the Bosonic Function, and $\widetilde{N}(\omega,T)$ is the normalized density of states. The real part of the self-energy, $\Sigma_1(\omega,T)$, can be calculated by using the Kramers-Kronig relations.

The normalized density of states $\widetilde{N}(\omega,T)$ is modeled by (29):

$$\widetilde{N}(\omega,T) = \begin{cases} \widetilde{N}(0,T) + [1-\widetilde{N}(0,T)](\omega/\Delta_{pg})^2 & |\omega| \leq \Delta_{pg} \\ 1 + 2/3[1-\widetilde{N}(0,T)] & |\omega| \in (\Delta_{pg}, 2\Delta_{pg}) \\ 1 & |\omega| \geq 2\Delta_{pg} \end{cases}$$

Where $\Delta_{pg}$ represents the (pseudo)gap width, while the normalized density of states at $E_F$, i.e., $\widetilde{N}(0,T)$, represents the gap filling. The $\Pi(\Omega)$ function is extracted by fitting the EDM to the normal state optical properties (25). $\Pi(\Omega)$ is characterized by a low-energy part (up to 40 meV), a peak centered at ~60 meV and a broad continuum extending up to 350 meV.

The analysis of the time- and frequency-resolved data is performed by modelling the non-equilibrium dielectric function ($\varepsilon_{neq}(\omega)$) and calculating the reflectivity variation through the expression: $\delta R/R(\omega,t)=[R_{neq}(\omega,t)-R_{eq}(\omega)]/R_{eq}(\omega)$, where the normal incidence reflectivities are calculated as: $R_{eq}(\omega)=|[1-\sqrt{\varepsilon_{eq}(\omega)}]/[1+\sqrt{\varepsilon_{eq}(\omega)}]|^2$ and $R_{neq}(\omega)=|[1-\sqrt{\varepsilon_{neq}(\omega)}]/[1+\sqrt{\varepsilon_{neq}(\omega)}]|^2$.

The role of the finite penetration depth of the pump pulse ($d_{pu}$=160 nm @ 1.55 eV) is accounted for by numerically calculating $\delta R/R(\omega)$ through a transfer matrix method, when a graded index of the variation of the refractive index $n$ with exponential profile along the direction $z$ perpendicular to the surface, i.e., $\delta n=\delta n_0\exp(-z/d_{pu})$, is assumed.

**Photoemission spectroscopy**
The Fermi surface of Y-Bi2212 reported in Fig. 4a has been measured by ARPES in equilibrium conditions. ARPES has been performed with 21.2 eV linearly polarized photons (He-α line from a SPECS UVS300 monochromatized lamp) and a SPECS Phoibos 150 hemispherical analyzer. Energy and angular resolutions were set to 30 meV and 0.2°. The Bi2212 samples studied by TR-ARPES are nearly optimally doped single crystals with a transition temperature $T_c$=88 K. The samples have been excited by 55 fs laser pulses with a photon energy of 1.55 eV at 300 kHz repetition rate, at an absorbed fluence of 35 μJ/cm². The transient electron distribution was probed by time-delayed 80 fs, 6 eV laser pulses, photoemitting electrons which were detected by a time-of-flight spectrometer. The energy resolution was 50 meV, the momentum resolution 0.05 Å⁻¹ and the time resolution <100 fs.

**Cluster-Dynamical Mean-Field Theory and the Hubbard model**
The Hubbard Hamiltonian (48,49) is given by:

$$\hat{H} = -\sum_{i,j,\sigma}(t_{i,j}\hat{c}^{\dagger}_{i\sigma}\hat{c}_{j\sigma} + \text{c.c.}) + U\sum_i \hat{n}_{i,\uparrow}\hat{n}_{i,\downarrow} - \mu\sum_i \hat{n}_i$$

where $\hat{c}^{\dagger}_{i\sigma}$ ($\hat{c}_{j\sigma}$) create (annihilate) an electron with spin σ on the $i$ ($j$) site, $\hat{n}_{i,\sigma} = \hat{c}^{\dagger}_{i\sigma}\hat{c}_{i\sigma}$ is the number operator, $t_{ij}$ the hopping amplitude to nearest and next-nearest neighbors, $U$ the Coulomb repulsion between two electrons occupying the same lattice site and $\mu$ is the chemical potential that controls the total number of electrons $n = \sum_{i\sigma} <\hat{n}_{i,\sigma}> / N$ in the $N$ sites.

The Hubbard model has been studied by means of a Cluster Dynamical Mean Field Theory that maps the full lattice model onto a finite small cluster (here a 4-site cluster) embedded in an effective medium which is self-consistently determined as in standard mean-field theory. The method therefore fully accounts for the short-range quantum correlations inside the cluster. It has been shown by various authors that different implementations of this approach provide qualitatively similar results and reproduce the main features of the phase diagram of the cuprates, including the d-wave superconducting state and the pseudogap region that we discuss in the present paper. The calculations of this paper use the Dynamical Cluster Approximation (59) prescription and the 4* patching of the Brillouin zone introduced in (16) and have been performed using finite temperature Exact Diagonalization (60) (ED) to solve the self-consistent cluster problem using 8 energy levels in the bath as in several previous calculations. The finite-temperature version of the ED has been implemented as discussed in (16) including typically 40 states in the low-temperature expansion of the observables.

**References:**

1. Lee, P. A., Nagaosa, N. & Wen, X.-G. Doping a Mott insulator: Physics of high-temperature superconductivity. *Rev. Mod. Phys.* **78**, 17 (2006).
2. Damascelli, A., Hussain, Z. & Shen, Z.-X. Angle-resolved photoemission studies of the cuprate superconductors. *Rev. Mod. Phys.* **75**, 473 (2003).
3. Varma, C. M. Theory of the pseudogap state of the cuprates. *Phys. Rev. B* **73**, 155113 (2006).
4. Sachdev, S. *Quantum Phase Transitions* (Cambridge University Press, 2011) [2nd Edition].
5. Fauqué, B., Sidis, Y., Hinkov, V., Pailhès, S., Lin, C. T., Chaud, X. & Bourges, P. Magnetic Order in the Pseudogap Phase of High-$T_C$ Superconductors. *Phys. Rev. Lett.* **96**, 197001 (2006).
6. Li, Y., Balédent, V., Barišić, N., Cho, Y., Fauqué, B., Sidis, Y., Yu, G., Zhao, X., Bourges, P. & Greven, M. Unusual magnetic order in the pseudogap region of the superconductor $HgBa_2CuO_{4+\delta}$. *Nature* **455**, 18 (2008).
7. Li, Y., Balédent, V., Yu, G., Barišić, N., Hradil, K., Mole, R. A., Sidis, Y., Steffens, P., Zhao, X., Bourges, P. & Greven, M. Hidden magnetic excitation in the pseudogap phase of a high-$T_C$ superconductor. *Nature* **468**, 283 (2010).
8. Ghiringhelli, G., Le Tacon, M., Minola, M., Blanco-Canosa, S., Mazzoli, C., Brookes, N. B., De Luca, G. M., Frano, A., Hawthorn, D. G., He, F., Loew, T., Moretti Sala, M., Peets, D. C., Salluzzo, M., Schierle, E., Sutarto, R., Sawatzky, G. A., Weschke, E., Keimer, B. & Braicovich, L. Long-Range Incommensurate Charge Fluctuations in $(Y,Nd)Ba_2Cu_3O_{6+x}$. *Science* **337**, 821 (2012).
9. Chang, Y., Blackburn, E., Holmes, A. T., Christensen, N. B., Larsen, J., Mesot, J., Liang, R., Bonn, D. A., Hardy, W. N., Watenphul, A., v. Zimmermann, M., Forgan,

E. M. & Hayden, S. M. Direct observation of competition between superconductivity and charge density wave order in $YBa_2Cu_3O_{6.67}$. *Nature Physics* **8**, 871 (2012).

10. Tranquada, J. M., Sternlieb, B. J., Axe, J. D., Nakamura, Y. & Uchida, S. Evidence for stripe correlations of spins and holes in copper oxide superconductors. *Nature* **375**, 561 (1995).
11. Kivelson, S. A., Fradkin, E. & Emery, V. J. Electronic liquid-crystal phases of a doped Mott insulator. *Nature* **393**, 550 (1998).
12. Mesaros, A., Fujita, K., Eisaki, H., Uchida, S., Davis, J. C., Sachdev, S., Zaanen, J., Lawler, M. J. & Kim, E.-A. Topological Defects Coupling Smectic Modulations to Intra–Unit-Cell Nematicity in Cuprates. *Science* **333**, 426 (2011).
13. Hinkov, V., Haug, D., Fauqué, B., Bourges, P., Sidis, Y., Ivanov, A., Bernhard, C., Lin, C. T. & Keimer, B. Electronic Liquid Crystal State in the High-Temperature Superconductor $YBa_2Cu_3O_{6.45}$. *Science* **319**, 597 (2008).
14. Shekhter, A., Ramshaw, B. J., Liang, R., Hardy, W. N., Bonn, D. A., Balakirev, F. F., McDonald, R. D., Betts, J. B., Riggs, S. C. & Migliori, A. Bounding the pseudogap with a line of phase transitions in $YBa_2Cu_3O_{6+\delta}$. *Nature* **498**, 75 (2013).
15. Ferrero, M., Cornaglia, P. S., De Leo, L., Parcollet, O., Kotliar, G. & Georges, A. Pseudogap opening and formation of Fermi arcs as an orbital-selective Mott transition in momentum space. *Phys. Rev. B* **80**, 064501 (2009).
16. Gull, E., Ferrero, M., Parcollet, O., Georges, A. & Millis, A. J. Momentum-space anisotropy and pseudogaps: A comparative cluster dynamical mean-field analysis of the doping-driven metal-insulator transition in the two-dimensional Hubbard model. *Phys. Rev. B* **82**, 155101 (2010).
17. Civelli, M., Capone, M., Georges, A., Haule, K., Parcollet, O., Stanescu, T. D. & Kotliar, G. Nodal-Antinodal Dichotomy and the Two Gaps of a Superconducting Doped Mott Insulator. *Phys. Rev. Lett.* **100**, 046402 (2008).
18. Gedik, N., Blake, P., Spitzer, R. C., Orenstein, J., Liang, R., Bonn, D. A. & Hardy, W. N. Single-quasiparticle stability and quasiparticle-pair decay in $YBa_2Cu_3O_{6.5}$. *Phys. Rev. B* **70**, 014504 (2004).
19. Orenstein, J. Ultrafast spectroscopy of quantum materials. *Phys. Today* **65**, 44 (2012).
20. He, R.-H., Hashimoto, M., Karapetyan, H., Koralek, J. D., Hinton, J. P., Testaud, J. P., Nathan, V., Yoshida, Y., Yao, H., Tanaka, K., Meevasana, W., Moore, R. G., Lu, D. H., Mo, S.-K., Ishikado, M., Eisaki, H., Hussain, Z., Devereaux, T. P., Kivelson, S. A., Orenstein, J., Kapitulnik, A. & Shen, Z.-X. From a Single-Band Metal to a High-Temperature Superconductor via Two Thermal Phase Transitions. *Science* **331**, 1579 (2011).
21. Liu, Y. H., Toda, Y., Shimatake, K., Momono, N., Oda, M. & Ido, M., Direct Observation of the Coexistence of the Pseudogap and Superconducting Quasiparticles in $Bi_2Sr_2CaCu_2O_{8+\delta}$ by Time-Resolved Optical Spectroscopy. *Phys. Rev. Lett.* **101**, 137003 (2008).
22. Demsar, J., Podobnik, B., Kabanov, V. V., Wolf, T. & Mihailovic, D. Superconducting Gap $\Delta_c$, the Pseudogap $\Delta_p$, and Pair Fluctuations above $T_c$ in

Overdoped $Y_{1-x}Ca_xBa_2Cu_3O_{7-\delta}$ from Femtosecond Time-Domain Spectroscopy. *Phys. Rev. Lett.* **82**, 4918 (1999).
23. Giannetti, C., Cilento, F., Dal Conte, S., Coslovich, G., Ferrini, G., Molegraaf, H., Raichle, M., Liang, R., Eisaki, H., Greven, M., Damascelli, A., van der Marel, D. & Parmigiani, F. Revealing the high-energy electronic excitations underlying the onset of high-temperature superconductivity in cuprates. *Nat. Commun.* **3**, 253 (2011).
24. Eichberger, M., Schafer, H., Krumova, M., Beyer, M., Demsar, J., Berger, H., Moriena, G., Sciaini, G. & Miller, R. J. D. Snapshots of cooperative atomic motions in the optical suppression of charge density waves. *Nature* **468**, 799 (2010).
25. Dal Conte, S., Giannetti, C., Coslovich, G., Cilento, F., Bossini, D., Abebaw, T., Banfi, F., Ferrini, G., Eisaki, H., Greven, M., Damascelli, A., van der Marel, D. & Parmigiani, F. Disentangling the Electronic and Phononic Glue in a High-$T_c$ Superconductor. *Science* **335**, 1600 (2012).
26. Basov, D.N. & Timusk, T. Electrodynamics of high-$T_c$ superconductors. Rev. Mod. Phys. **77**, 721 (2005).
27. van Heumen, E., Muhlethaler, E., Kuzmenko, A.B., Eisaki, H., Meevasana, W., Greven, M., and van der Marel, D. Optical determination of the relation between the electron-boson coupling function and the critical temperature in high-$T_c$ cuprates. *Phys. Rev. B* **79**, 184512 (2009).
28. Sharapov, S. G. & Carbotte, J. P. Effects of energy dependence in the quasiparticle density of states on far-infrared absorption in the pseudogap state. *Phys. Rev. B* **72**, 134506 (2005).
29. Hwang, J., Electron-boson spectral density function of underdoped $Bi_2Sr_2CaCu_2O_{8+\delta}$ and $YBa_2Cu_3O_{6.5}$. *Phys. Rev. B* **83**, 014507 (2011).
30. Cilento, F., Dal Conte, S., Coslovich, G., Banfi, F., Ferrini, G., Eisaki, H., Greven, M., Damascelli, A., van der Marel, D., Parmigiani, F. & Giannetti, C. In search for the pairing glue in cuprates by non-equilibrium optical spectroscopy. *J. Phys.: Conf. Ser.* **449**, 012003 (2013).
31. Renner, Ch., Revaz, B., Genoud, J.-Y., Kadowaki, K. & Fischer, O. Pseudogap Precursor of the Superconducting Gap in Under- and Overdoped $Bi_2Sr_2CaCu_2O_{8+\delta}$. *Phys. Rev. Lett.* **80**, 149 (1998).
32. Smallwood, C. L., Zhang, W., Miller, T. L., Jozwiak, C., Eisaki, H., Lee, D.-H. & Lanzara, A. Time- and momentum-resolved gap dynamics in $Bi_2Sr_2CaCu_2O_{8+\delta}$, *Phys. Rev. B* **89**, 115126 (2014).
33. Mirzaei, S. I., Stricker, D., Hancock, J. N., Berthod, C., Georges, A., van Heumen, E., Chan, M. K., Zhao, X., Li, Y., Greven, M., Barišić, N. & van der Marel, D. Spectroscopic evidence for Fermi liquid-like energy and temperature dependence of the relaxation rate in the pseudogap phase of the cuprates. *Proc. Natl. Acad. Sci.* **110**, 5774 (2013).
34. Barišić, N., Chan, M. K., Li, Y., Yu, G., Zhao, X., Dressel, M., Smontara, A. & Greven, M. Universal sheet resistance and revised phase diagram of the cuprate high-temperature superconductors. *Proc. Natl. Acad. Sci.* **110**, 12235 (2013).

35. Cortés, R., Rettig, L., Yoshida, Y., Eisaki, H., Wolf, M. & Bovensiepen, U. Momentum-Resolved Ultrafast Electron Dynamics in Superconducting $Bi_2Sr_2CaCu_2O_{8+\delta}$, *Phys. Rev. Lett.* **107**, 097002 (2011).
36. Smallwood, C. L., Hinton, J. P., Jozwiak, C., Zhang, W., Koralek, J. K., Eisaki, H., Lee, D. H., Orenstein, J. & Lanzara, A. Tracking Cooper Pairs in a Cuprate Superconductor by Ultrafast Angle-Resolved Photoemission. *Science* **336**, 1137 (2012).
37. Eisaki, H., Kaneko, N., Feng, D. L., Damascelli, A., Mang, P. K., Shen, K. M., Shen, Z.-X. & Greven, M. Effect of chemical inhomogeneity in bismuth-based copper oxide superconductors. *Phys. Rev. B* **69**, 064512 (2004).
38. Barišić, N., Li, Y., Zhao, X., Cho, Y.-C., Chabot-Couture, G., Yu, G. & Greven, M. Demonstrating the model nature of the high-temperature superconductor $HgBa_2CuO_{4+\delta}$. *Phys. Rev. B* **78**, 054518 (2008).
39. Deutscher, G., Santander-Syro, A. F. & Bontemps, N. Kinetic energy change with doping upon superfluid condensation in high-temperature superconductors. *Phys. Rev. B* **72**, 092504 (2005).
40. Gedik, N., Langner, M., Orenstein, J., Ono, S., Abe, Y. & Ando, Y. Abrupt Transition in Quasiparticle Dynamics at Optimal Doping in a Cuprate Superconductor System. *Phys. Rev. Lett.* **95**, 117005 (2005).
41. Carbone, F., Kuzmenko, A. B., Molegraaf, H. J. A., van Heumen, E., Lukovac, V., Marsiglio, F., van der Marel, D., Haule, K., Kotliar, G., Berger, H., Courjault, S., Kes, P. H. & Li, M., Doping dependence of the redistribution of optical spectral weight in $Bi_2Sr_2CaCu_2O_{8+\delta}$. *Phys. Rev. B* **74**, 064510 (2006).
42. Gull, E. & Millis, A. J. Energetics of superconductivity in the two-dimensional Hubbard model. *Phys. Rev. B* **86**, 241106(R) (2012).
43. De Almeida-Didry, S., Sidis, Y., Baledent, V., Giovannelli, F., Monot-Laffez, I. & Bourges, P. Evidence for intra-unit-cell magnetic order in $Bi_2Sr_2CaCu_2O_{8+\delta}$. *Phys. Rev. B* **86**, 020504(R) (2012).
44. Raffy, H., Toma, V., Murrills, C., Li, Z. Z. *c*-axis resistivity of $Bi_2Sr_2CaCu_2O_y$ thin films at various oxygen doping: Phase diagram and scaling law. *Physica C* **460**, 851 (2007).
45. Xia, J., Schemm, E., Deutscher, G., Kivelson, S. A., Bonn, D. A., Hardy, W. N., Liang, R., Siemons, W., Koster, G., Fejer, M. M. & Kapitulnik, A. Polar Kerr-Effect Measurements of the High-Temperature $YBa_2Cu_3O_{6+x}$ Superconductor: Evidence for Broken Symmetry near the Pseudogap Temperature. *Phys. Rev. Lett.* **100**, 127002 (2008).
46. Blackburn, E., Chang, J., Hücker, M., Holmes, A. T., Christensen, N. B., Liang, R., Bonn, D. A., Hardy, W. N., Rütt, U., Gutowski, O., Zimmermann, M. v., Forgan, E. M. & Hayden, S. M. X-Ray Diffraction Observations of a Charge-Density-Wave Order in Superconducting Ortho-II $YBa_2Cu_3O_{6.54}$ Single Crystals in Zero Magnetic Field. *Phys. Rev. Lett.* **110**, 137004 (2013).

47. Parker, C. V., Aynajian, P., da Silva Neto, E. H., Pushp, A., Ono, S., Wen, J., Xu, Z., Gu, G. & Yazdani, A. Fluctuating stripes at the onset of the pseudogap in the high-$T_c$ superconductor $Bi_2Sr_2CaCu_2O_{8+x}$. *Nature* **468**, 677 (2010).
48. Hubbard, J. Electron Correlations in Narrow Energy Bands. *Proc. R. Soc. Lond. A* **276**, 238 (1963).
49. Anderson, P. W. Theory of Magnetic Exchange Interactions: Exchange in Insulators and Semiconductors. *Solid State Phys.* **14**, 99 (1963).
50. Sordi, G., Haule, K. & Tremblay, A.-M. S. Mott physics and first-order transition between two metals in the normal-state phase diagram of the two-dimensional Hubbard model. *Phys. Rev. B* **84**, 075161 (2011).
51. Yusupov, R., Mertelj, T., Kabanov, V.V., Brazovskii, S., Kusar, P., Chu, J.-H., Fisher, I.R. and Mihailovic, D. Coherent dynamics of macroscopic electronic order through a symmetry breaking transition. *Nature Physics* **6**, 681 (2010).
52. Castellani, C., Di Castro, C. & Grilli, M. Singular Quasiparticle Scattering in the Proximity of Charge Instabilities. *Phys. Rev. Lett.* **75**, 4650 (1995).
53. Comin, R. *et al*. Charge order driven by Fermi-arc instability in $Bi_2Sr_{2-x}La_xCuO_{6+\delta}$. Accepted for publication on *Science*.
54. Rosen, J. A., Comin, R., Levy, G., Fournier, D., Zhu, Z.-H., Ludbrook, B., Veenstra, C. N., Nicolaou, A., Wong, D., Dosanjh, P., Yoshida, Y., Eisaki, H., Blake, G. R., White, F., Palstra, T. T. M., Sutarto, R., He, F., Frano, A., Lu, Y., Keimer, B., Sawatzky, G. A., Petaccia, L. & Damascelli, A. Surface-enhanced charge-density-wave instability in underdoped Bi2201. *Nat. Commun.* **4**, 1977 (2013).
55. Hirsch, J. E.. Charge expulsion, charge inhomogeneity and phase separation in dynamic Hubbard models. Preprint at http://arxiv.org/abs/1307.6526 (2013).
56. Kaiser, S., Nicoletti, D., Hunt, C.R., Hu, W., Gierz, I., Liu, H.Y., Le Tacon, M., Loew, T., Haug, D., Keimer, B., & Cavalleri, A.. Light-induced inhomogeneous superconductivity far above $T_c$ in $YBa_2Cu_3O_{6+x}$. Preprint at http://arxiv.org/abs/1205.4661 (2012).
57. van der Marel, D., Molegraaf, H. J. A., Zaanen, J., Nussinov, Z., Carbone, F., Damascelli, A., Eisaki, H., Greven, M., Kes, P. H. & Li, M. Quantum critical behaviour in a high-$T_c$ superconductor. *Nature* **425**, 271 (2003).
58. Cilento, F., Giannetti, C., Ferrini, G., Dal Conte, S., Sala, T., Coslovich, G., Rini, M., Cavalleri, A. & Parmigiani, F. Ultrafast insulator-to-metal phase transition as a switch to measure the spectrogram of a supercontinuum light pulse. *Appl. Phys. Lett.* **96**, 021102 (2010).
59. Maier, T., Jarrell, M., Pruschke, T. & Hettler, M. H. Quantum cluster theories. *Rev. Mod. Phys.* **77**, 1027 (2005).
60. Capone, M., de' Medici, L. & Georges, A. Solving the dynamical mean-field theory at very low temperatures using the Lanczos exact diagonalization. *Phys. Rev. B* **76**, 245116 (2007).

**Acknowledgments:**

The research activities of C.G., S.D.C., N.M., S.M., F.B., G.F., M.C., F.P. and U.B. have received funding from the European Union, Seventh Framework Programme (FP7 2007-2013), under Grant No. 280555 (GO FAST). F.C., G.C., and F.P. acknowledge the support of the Italian Ministry of University and Research under Grant No. FIRB-RBAP045JF2 and FIRB-RBAP06AWK3. M.C. is financed by European Research Council through FP7/ERC Starting Grant SUPERBAD, Grant Agreement 240524. The YBi2212 crystal growth work was performed in M.G.'s prior laboratory at Stanford University, Stanford, CA 94305, USA, and supported by DOE-BES. The Hg1201 crystal growth work at the University of Minnesota was supported by DOE-BES. The work at UBC was supported by the Killam, Sloan Foundation, CRC, and NSERCs Steacie Fellowship Programs (A. D.), NSERC, CFI, CIFAR Quantum Materials, and BCSI. D. v.d.M. acknowledges the support of the Swiss National Science Foundation under Grant No. 200020-140761 and MaNEP.

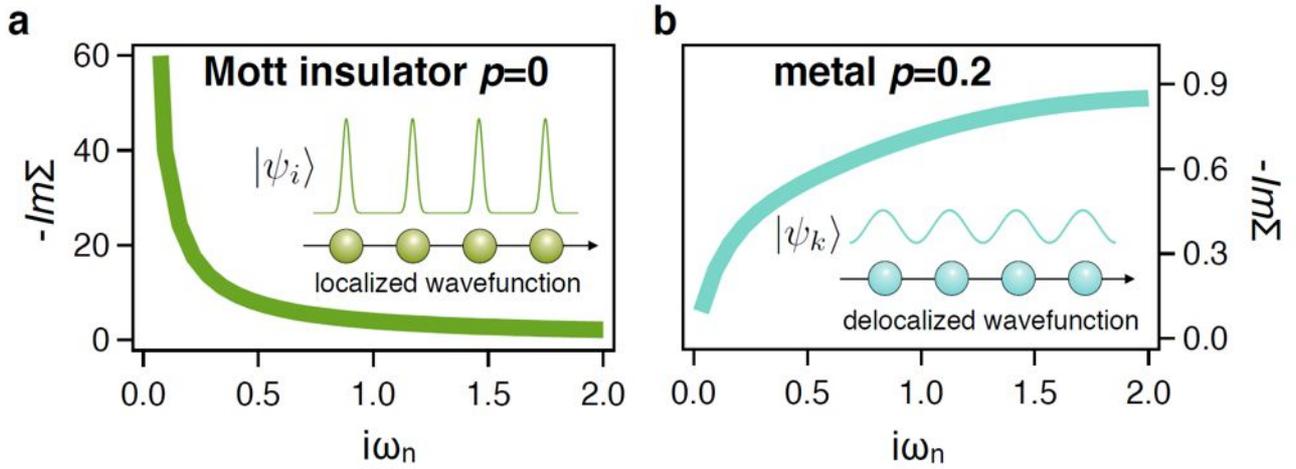

**Figure 1 | Electronic scattering rate of correlated systems**. **a)** and **b)** The imaginary part of the electronic self energy is calculated, as a function of the Matsubara frequency $w_n=(2n+1)\pi k_B T$, by solving the single-band Hubbard model via single-site Dynamical Mean Field Theory. The prototypical cases of the Mott insulator ($p=0$) and metallic system ($p=0.2$) are reported and contrasted. In the more realistic three-bands Hubbard model the same divergence of the self-energy is obtained as the charge-transfer insulating state is approached.

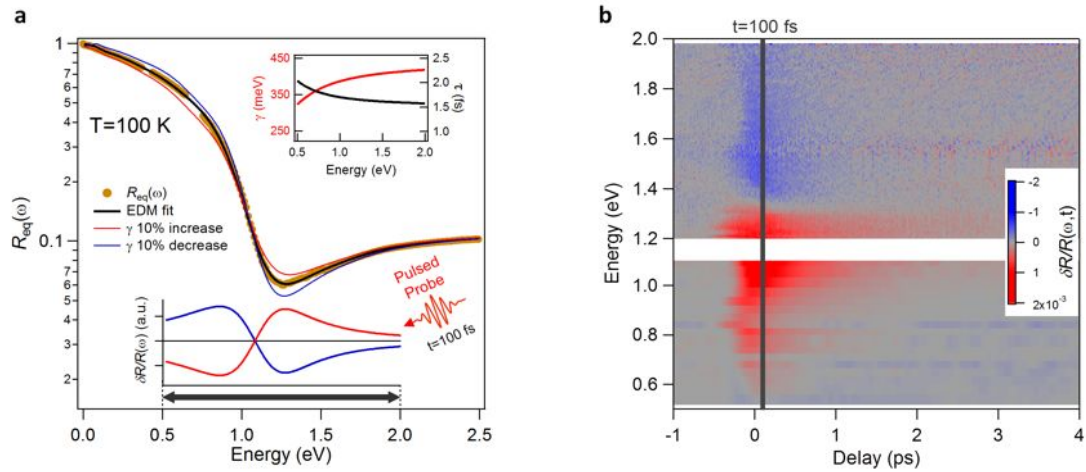

**Figure 2 | Equilibrium and non-equilibrium optical spectroscopies. a**) The equilibrium reflectivity of an OP YBi2212 sample, measured at *T*=100 K in the 0-2.5 eV spectral range, is reported (yellow dots). The black line is the fit of the equilibrium Extended Drude Model (EDM) to the data. The red (blue) curve is the reflectivity obtained from the EDM, in which an increase (decrease) of the optical scattering rate is artificially introduced. The bottom panel displays the relative reflectivity variation in the case of an increase (red line) and a decrease (blue line) of the optical scattering rate. The optical scattering rate, $\gamma(\omega) = \hbar/\tau(\omega) = \hbar\omega_p^2/4\pi \, \text{Re}(1/\sigma_D(\omega))$, is reported in the inset. The grey bar highlights the spectral range probed by the non-equilibrium spectroscopy. **b**) The relative reflectivity variation, $\delta R/R(\omega,t)$, is reported as a function of the probe photon energy ($\hbar\omega$) and pump-probe delay t, in the form of an intensity map. The color coding for $\delta R/R(\omega,t)$ is indicated by the colorscale.

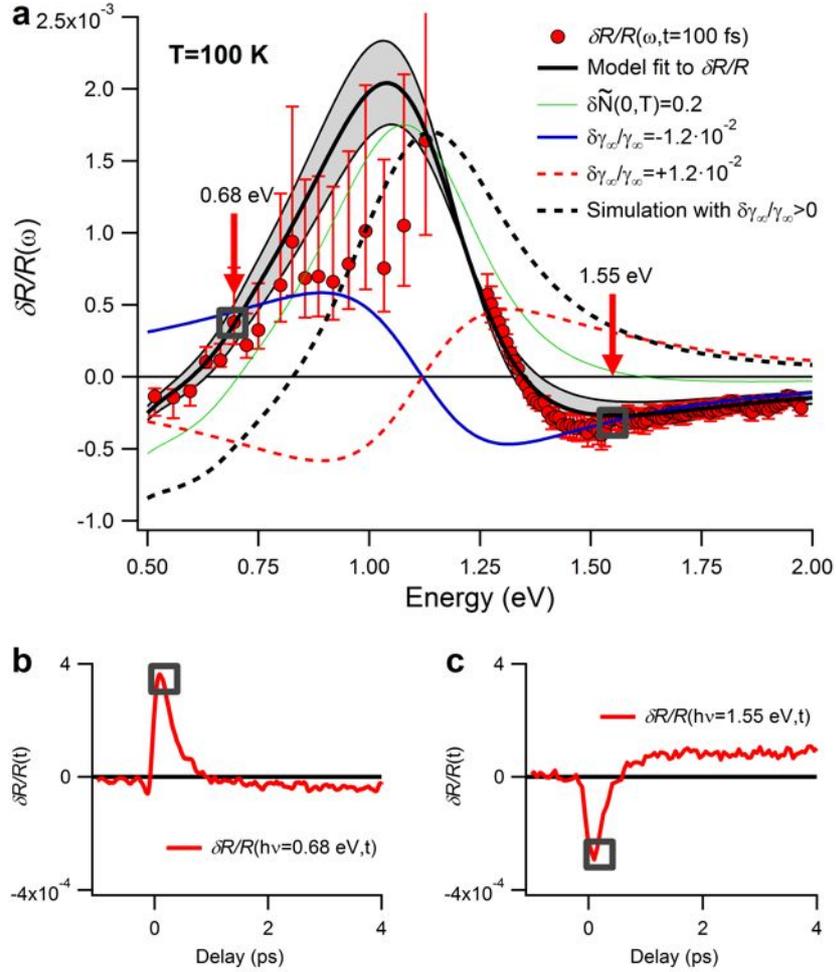

**Figure 3 | Differential model for the non-equilibrium reflectivity. (a)** The δR/R(ω, t=100 fs), measured on the OP-YBi2212 sample at T=100 K, is reported (red dots). The solid black line is the result of the fitting procedure of the differential model to δR/R(ω, t=100 fs), in which the density of states $\tilde{N}(\omega,T)$ (see Methods) and $\gamma_\infty$ have been considered as free parameters. The contribution of each effect is shown separately: the green line represents the filling of the pseudogap density of states ($\delta\tilde{N}\equiv\tilde{N}_{neq}(0,T)-\tilde{N}_{eq}(0,T)=0.2\pm0.02$), in agreement with recent time-resolved ARPES results (32); the blue line is related to the relative decrease of $\gamma_\infty$. The dashed red line represents the δR/R(ω) due to an increase of the scattering rate. The dashed black line represents the calculation of δR/R(ω), assuming $\delta\tilde{N}=0.2$ and $\delta\gamma_\infty/\gamma_\infty=+1.2\cdot10^{-2}$. The red arrows highlight the probe energies at which the dynamics of the $\delta\gamma_\infty/\gamma_\infty$ variation is investigated. The measurements at ω<1.2 eV have been performed by tuning the probe wavelength through an OPA (see Methods). Therefore each point at ω<1.2 eV belongs to a different time-domain measurement in which the uncertainty in the absolute amplitude (reported as a red bar) is larger than that of the measurements in the ω>1.2 eV frequency range. The error bars, accounting for the uncertainties in the pump fluence, size and spatial overlap with the probe beam, have been used as the weight for the fit. **b)** and **c)** Single-color time-traces at photon energies $\hbar\omega$=0.68 eV and $\hbar\omega$=1.55 eV.

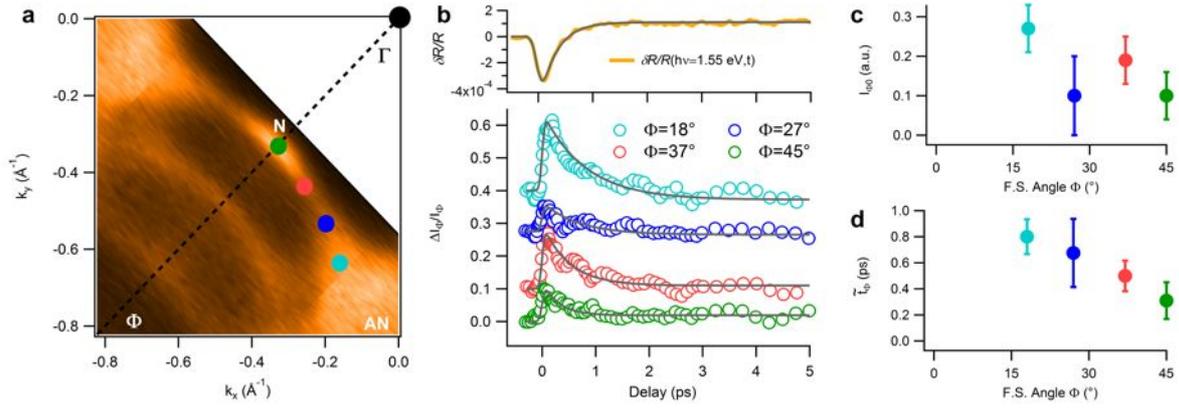

**Figure 4 | Equilibrium and non-equilibrium photoemission.** (**a**) The YBi2212 Fermi surface (FS), measured by conventional ARPES at 100 K, is reported. The two FS replicas surrounding the main FS are due to surface reconstruction. The Γ point and the FS **k**-space regions where the TR-ARPES measurements have been performed are indicated by full dots. Φ is the FS angle from the antinode. (**b**) The normalized integrated variation of the TR-ARPES intensity above $E_F$, $\Delta I_\Phi(t)/I_\Phi$, is reported for different points along the Fermi arcs (Φ=18°, 27°, 37°, 45°). Solid lines are the fit to the data with a single exponential decay, $\Delta I_{\Phi 0} \exp(-t/\tilde{t})$. In the top panel, the $\delta R/R(t)$ trace at 1.55 eV is reported for comparison. (**c** and **d**) The maximum intensity variation, $\Delta I_{\Phi 0}$, and the decay time, $\tilde{t}_\Phi$, of the non-equilibrium transient population measured by TR-ARPES are reported as a function of Φ. The error bars include the experimental uncertainty related to the possible alignment errors of the sample angle.

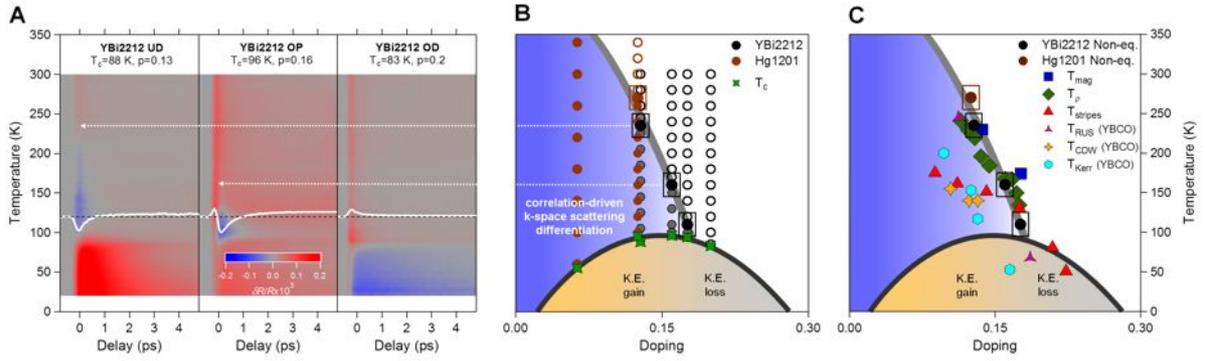

**Figure 5 | The cuprate phase diagram from non-equilibrium spectroscopy. (a)** The relative reflectivity variation $\delta R/R(t)$ measured at the probe energy of 1.55 eV, is reported as a function of the temperature (from 300 K to 20 K) for three YBi2212 samples with different hole concentrations: underdoped (UD, $T_c$=88 K), optimally doped (OP, $T_c$=96 K) and overdoped (OD, $T_c$=83 K). The white lines are the $\delta R/R(t)$ time-traces at 110 K. **(b)** The general phase diagram of cuprates, as unveiled by non-equilibrium reflectivity measurements, is sketched. The pseudogap boundary $T^*_{neq}$ (gray curve) is determined reporting the temperature at which a negative component in the $\delta R/R(t)$ signal appears on YBi2212 (black dots) and underdoped Hg1201 (purple dots) samples with $T_c$=55 K and $T_c$=95 K. The grey circles indicate some of the temperatures at which the $\delta R/R(t)$ data have been taken. The empty (full) circles corresponds to a (non) zero negative signal in the $\delta R/R(t)$ time-traces. The green markers denote the critical temperature $T_c$ of the samples. **(c)** The $T^*_{neq}(p)$ temperatures extracted from time-resolved optical spectroscopy experiments are compared to the values of $T^*(p)$ extracted from resistivity (44)(green diamonds) and ultrasound spectroscopy (14)(purple stars) on Bi2212 and YBCO and to the onset of the newly-discovered q=0 exotic magnetic modes (7,43) (blue squares) on Bi2212 and Hg1201. We also compare $T^*_{neq}(p)$ to the onset of ordered states, like CDW order (46)(yellow diamonds) in YBCO, fluctuating stripes (47)(red triangles) in Bi2212 and time-reversal symmetry breaking states in YBCO (45)(blue hexagons).

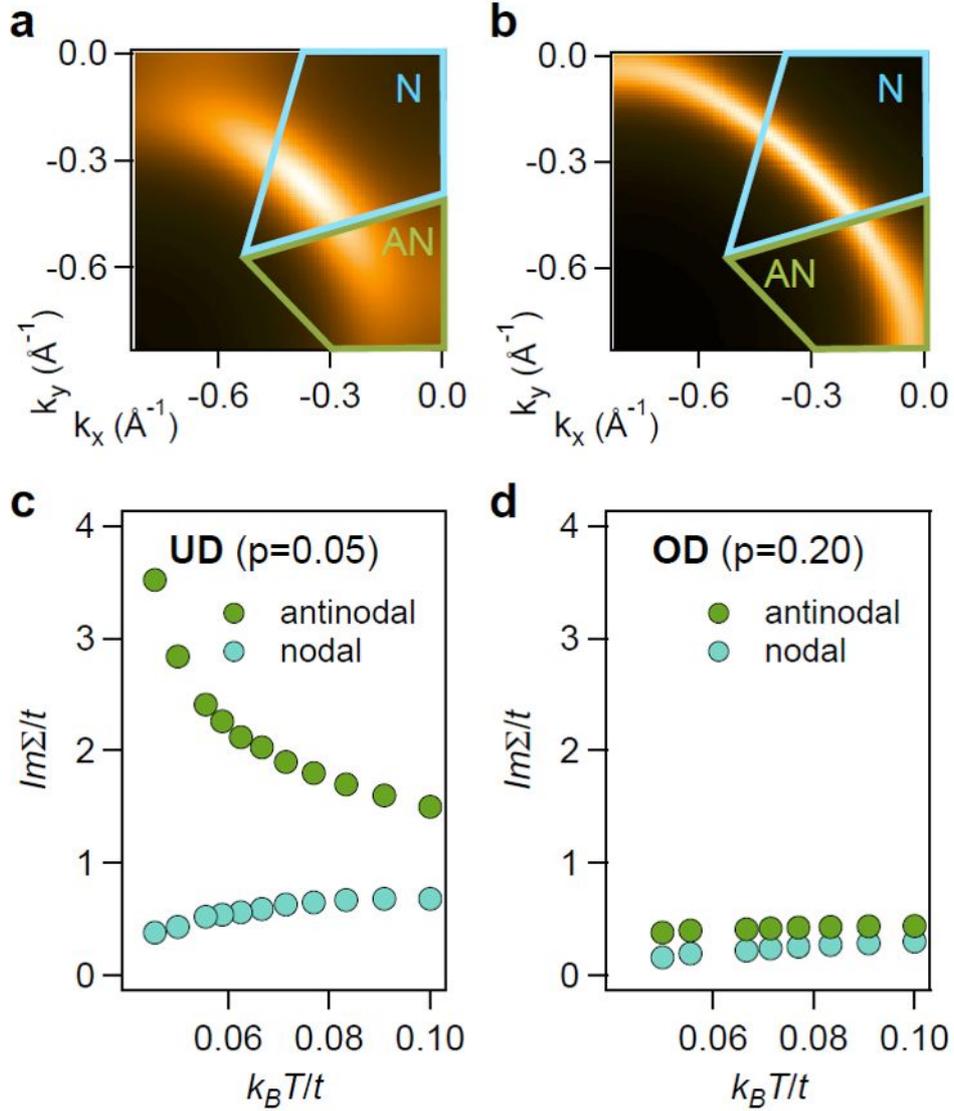

**Figure 6 | The single-band Hubbard model and CDMFT**. (**a** and **b**) The FS reconstructed from CDMFT, for an UD (*p*=0.05) and an OD (*p*=0.2) sample, are reported. (**c** and **d**) The imaginary parts of the electronic self-energy, calculated through CDMFT as a function of the temperature, are reported for the UD and the OD samples, respectively. The effective temperature is expressed in units of *t*. Assuming a reasonable value *t*=0.3 eV, *kT/t*=0.1 corresponds to *T*=350 K. The values of the parameters used in the calculations are *U*=9*t* and *t'*=-0.25*t*. The imaginary parts of the self-energies are calculated using DCA (Ref. 4 and Methods section) and 4 equal-area partitions of the Brillouin zone. Following Ref. 4, the nodal and antinodal self-energies are calculated in the regions centered in **k**=(0,0) (N) and **k**=(±π,0), (0,±π) (AN), as shown in panels **a** and **b**.

# Supplementary Material for:

*Photo-enhanced antinodal conductivity in the pseudogap state of high-$T_c$ cuprates*

**Contents**

**Supplementary Figures**

**Supplementary Notes**

**Supplementary References**

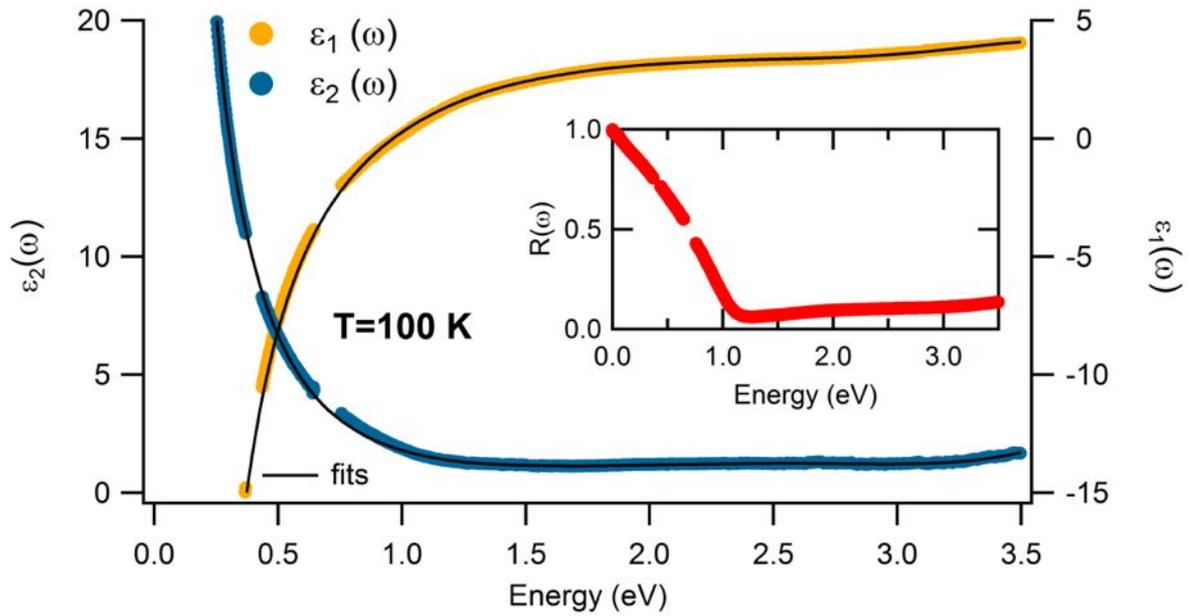

**Supplementary Figure 1 | Equilibrium Optical Properties at *T*=100 K.** The real part ($\varepsilon_1(\omega)$, right axis) and the imaginary part ($\varepsilon_2(\omega)$, left axis) of the equilibrium dielectric function of the optimally doped Y-Bi2212 sample, measured at *T*=100 K over a broad energy range, are reported. $R_{eq}(\omega)$ is reported in the inset. The dressed plasma frequency $\omega_p$ (defined by $\varepsilon_1(\omega_p)$=0) is 1.15 eV. The black curves are the best Extended Drude Model fits to the data.

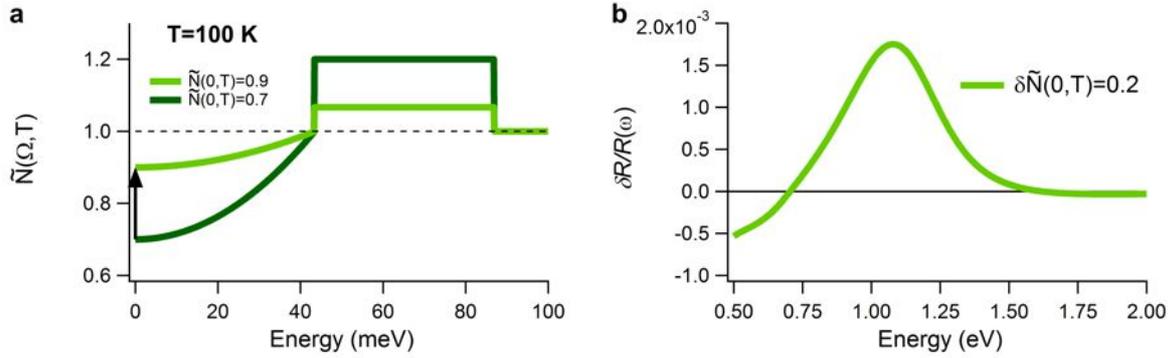

**Supplementary Figure 2 | Pseudogap Function and Photoinduced gap filling. a)** The pseudogap density of electronic states, $\widetilde{N}(\Omega,T)$, obtained by fitting the equilibrium dielectric function at $T$=100 K (see also Methods), is shown as a dark green line. The transient filling of the pseudogap (light green line) is simulated by increasing the normalized density of states at $E_F$, i.e., $\widetilde{N}(0,T)$. **b)** The contribution to the $\delta R/R(\omega,t)$, calculated through the differential Extended Drude Model in the case of $\delta\widetilde{N}(0,T)$=0.2, is shown.

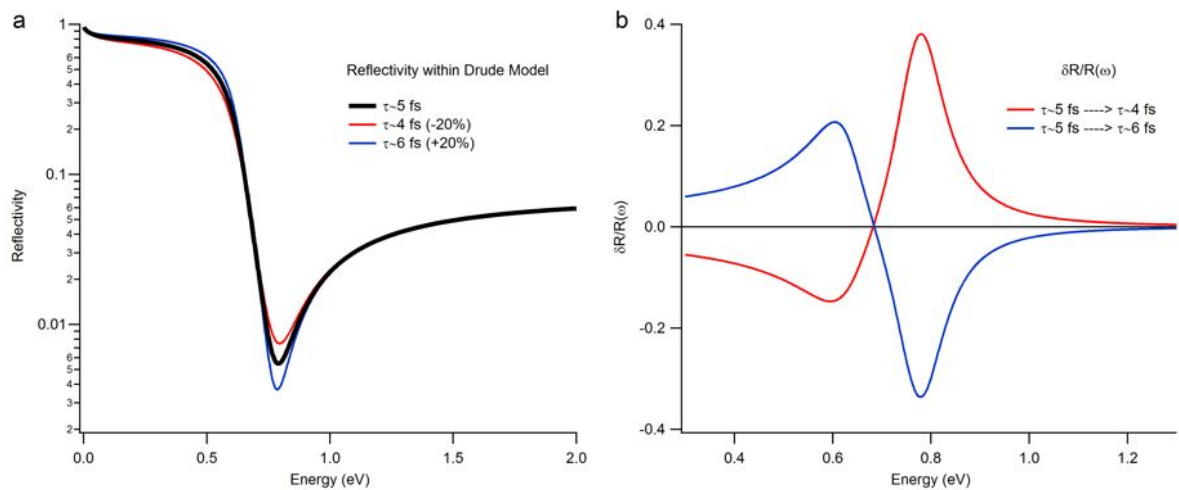

**Supplementary Figure 3 | Electronic Scattering Rate and Optical Properties. a.** The effect on the Reflectivity $R(\omega)$ of a ±20% modification of the electronic scattering time is shown. The plasma edge narrows (broadens), respectively. **b.** The same effect is shown in the $\delta R/R(\omega,t)$ signal.

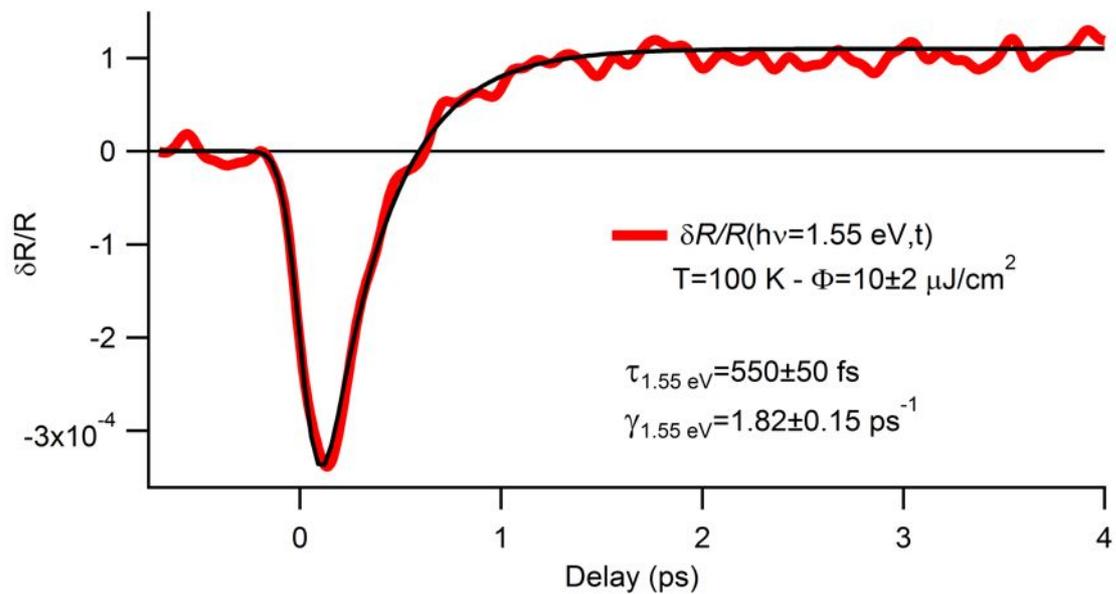

**Supplementary Figure 4 | Decay Time of the Non-Equilibrium Population.** The relaxation time of the δR/R(t) signal measured at 100 K, is reported.

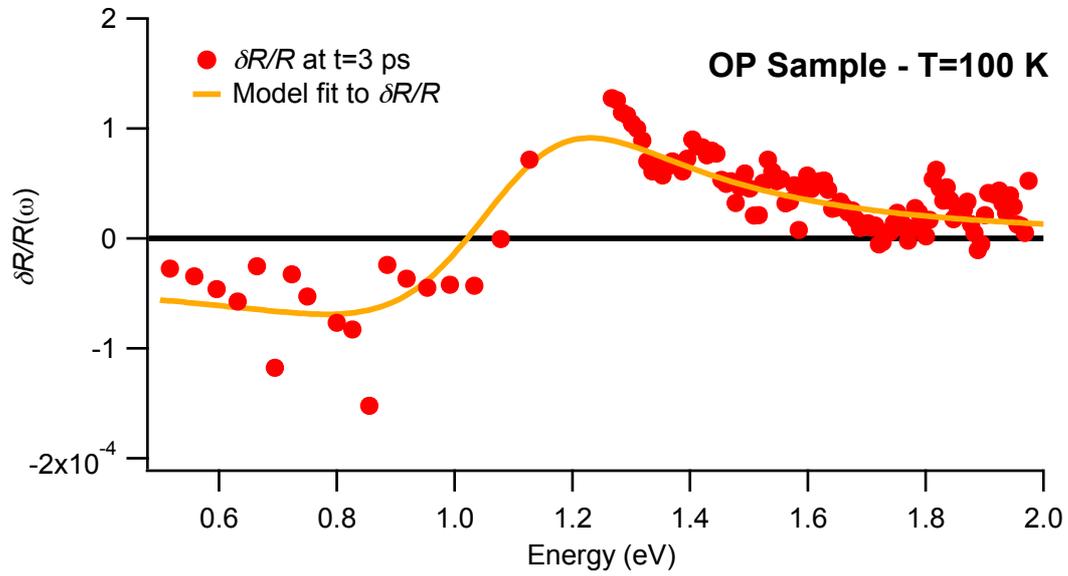

**Supplementary Figure 5 | Long-time reflectivity variation.** The $\delta R/R(\omega,t)$ signal measured at T=100 K on OP-YBi2212 is reported (red dots). The yellow line is the best fit to the data obtained by the procedure described in the text.

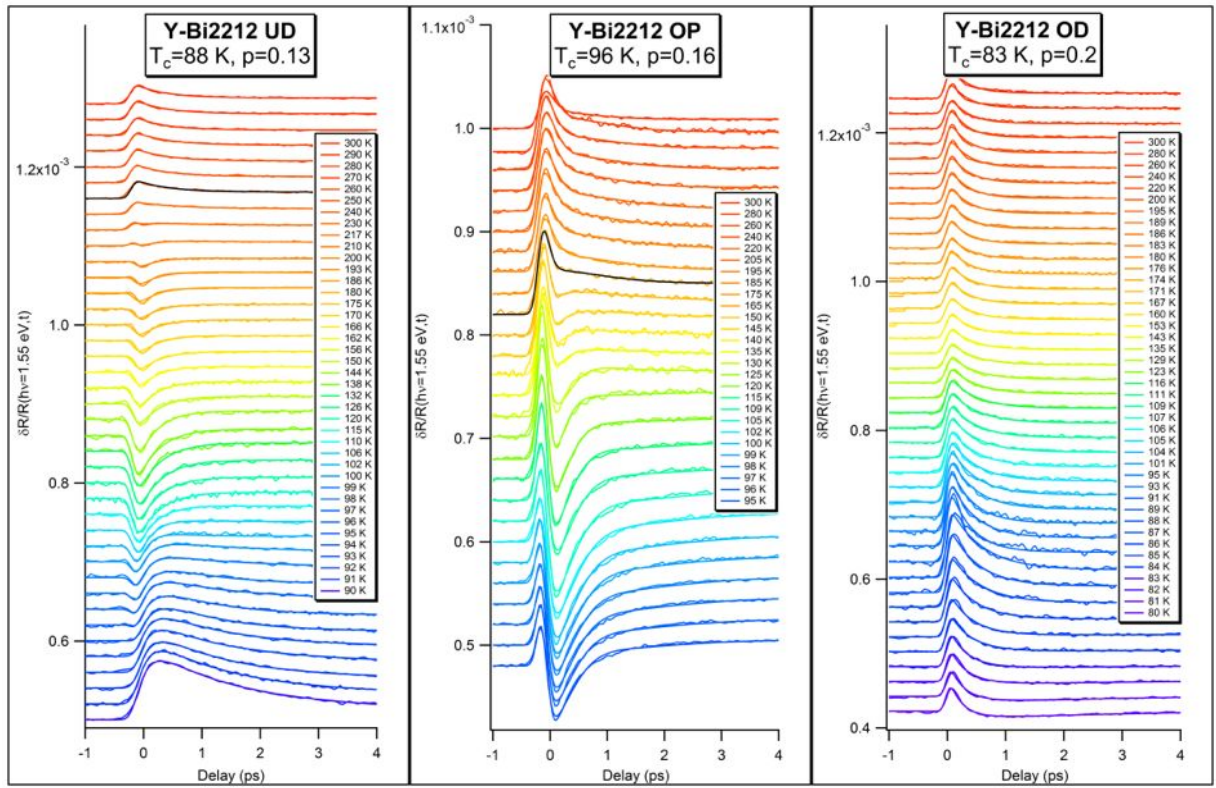

**Supplementary Figure 6 | Single-colour pump-probe measurements.** High-resolution, single-colour (1.55 eV probe photon energy) reflectivity measurements as a function of temperature are reported for three different samples: UD ($p$=0.13), OP ($p$=0.16), OD ($p$=0.2). The results of the fitting procedure are superimposed. Black fits are drawn in correspondence of the temperature below which the finite negative component in $f$(t) ($I_3$<0, see Supplementary Note 6) is observed.

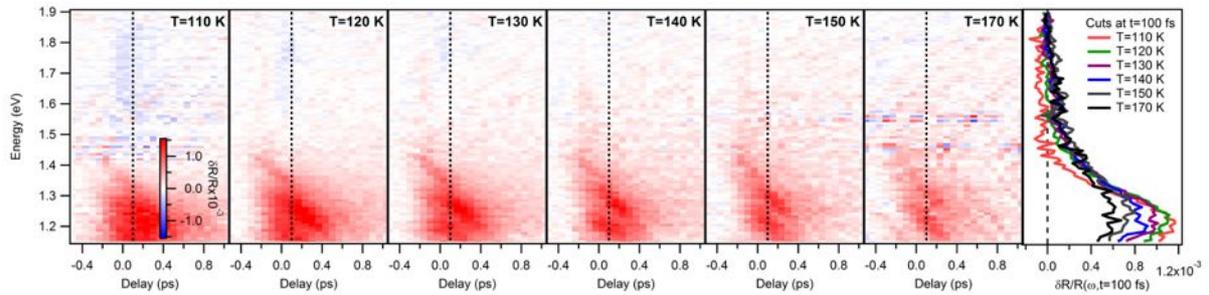

**Supplementary Figure 7 | Time-resolved optical spectroscopy on the sOD sample.** $\delta R/R(\omega,t)$ on the sOD sample is reported for some selected temperatures. Slices of the $\delta R/R(\omega,t)$ maps at t=100 fs are shown on the right panel.

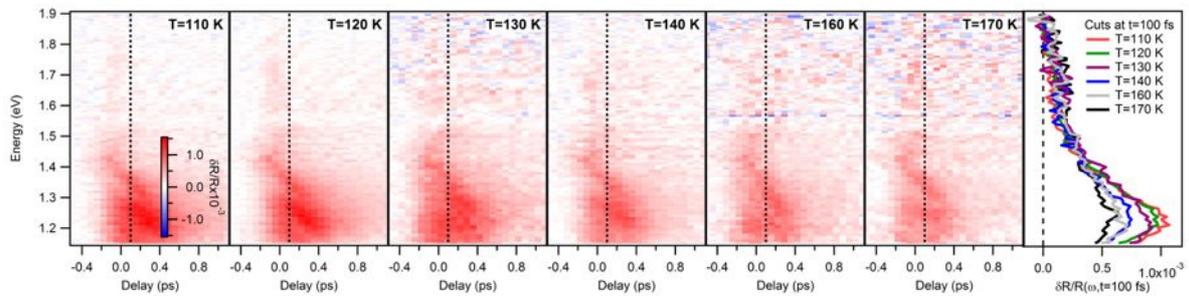

**Supplementary Figure 8 | Time-resolved optical spectroscopy on the OD sample.** $\delta R/R(\omega,t)$ on the OD sample is reported for some selected temperatures. Slices of the $\delta R/R(\omega,t)$ maps at t=100 fs are shown on the right panel.

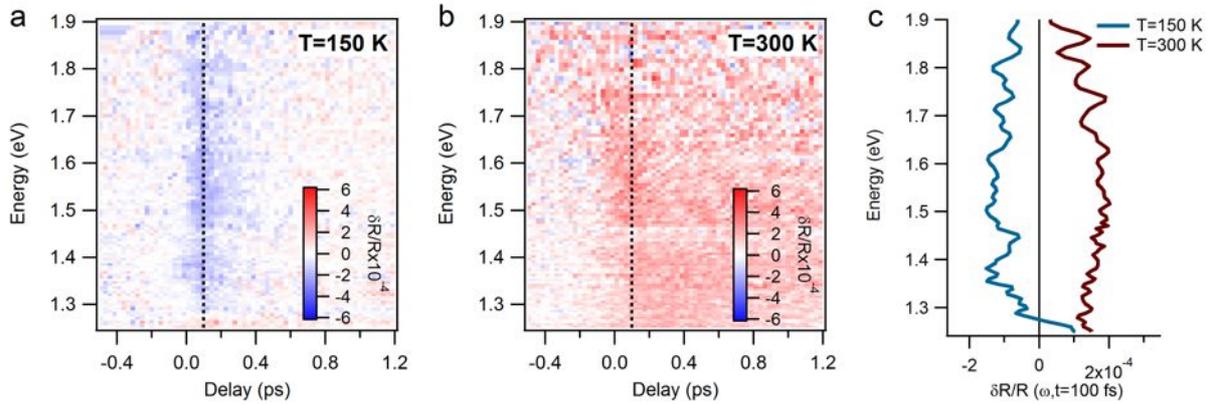

**Supplementary Figure 9 | Time resolved optical spectroscopy on sUD-Hg1201. a.** and **b.** The $\delta R/R(\omega,t)$ data, acquired respectively at $T$=150 K and $T$=300 K on the sUD-Hg1201 sample ($T_c$=95 K), are shown. **c.** The $\delta R/R(\omega)$ slices at t=100 fs are reported.

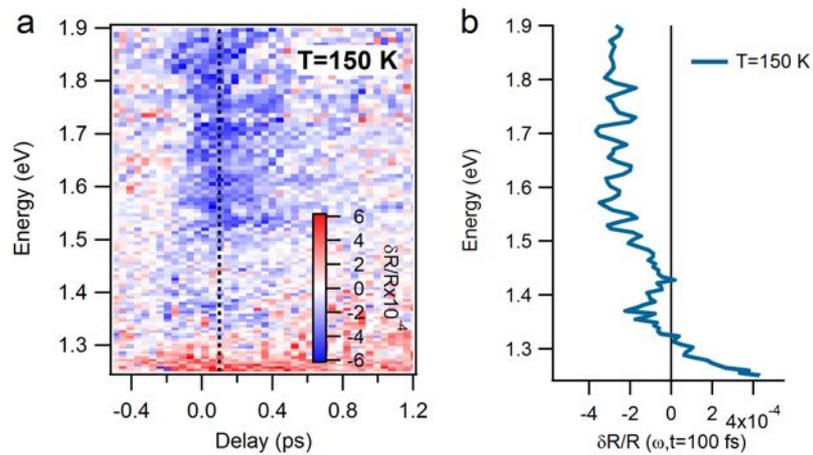

**Supplementary Figure 10 | Time resolved Optical Spectroscopic Measurements on UD Hg1201. a.** The $\delta R/R(\omega,t)$ data acquired at $T$=150 K on the UD Hg1201 sample ($T_c$=55 K) are shown. **b.** The $\delta R/R(\omega)$ slice at t=100 fs is reported.

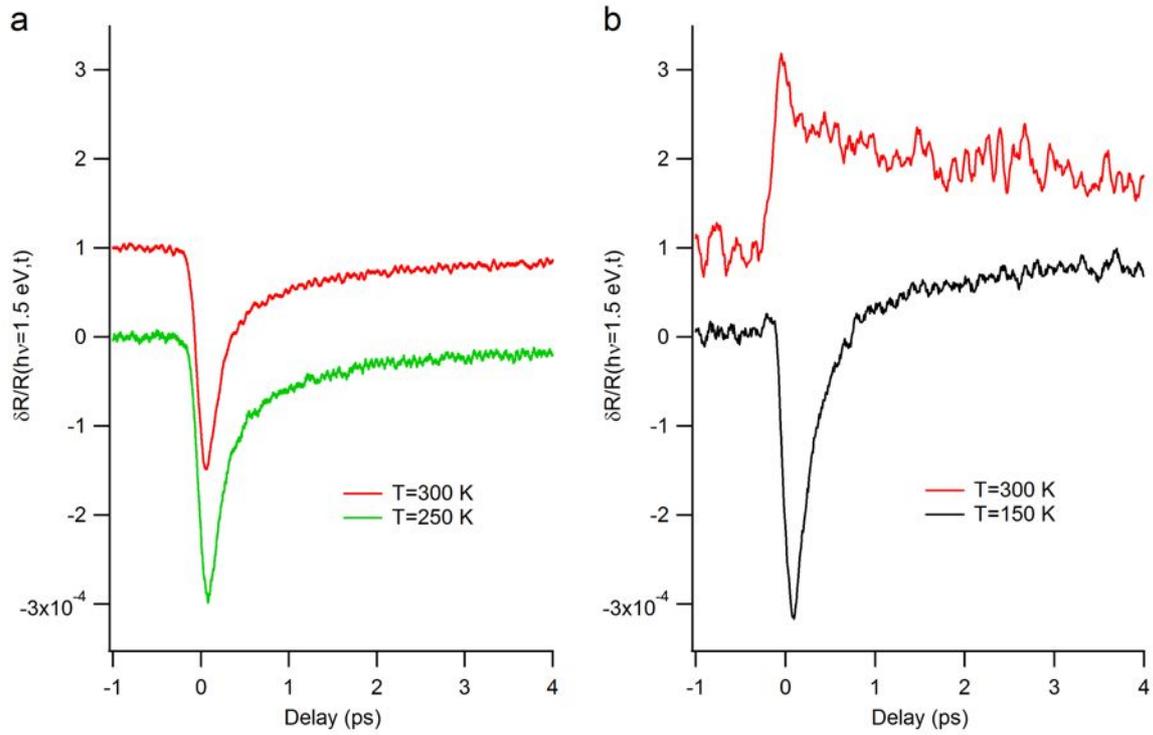

**Supplementary Figure 11 | Onset of the Pseudogap for Hg1201. a.** The high-resolution, single colour (1.55 eV probe energy) pump-probe measurements, performed on the UD-Hg1201, are shown. **b.** The same, but for sUD-Hg1201.

## Supplementary Note 1

### The Equilibrium Reflectivity at *T*=100 K

In Supplementary Fig. S1 we report the real and imaginary parts of the equilibrium dielectric function $\varepsilon_{eq}(\omega)$, measured by conventional reflectivity and ellipsometry techniques (1), in a wide spectral range (0-3.5 eV). The data reported refer to the *ab*-plane pseudo-dielectric function, which has been corrected for the *c*-axis admixture due to the off-normal angle of incidence used with ellipsometry.

A satisfactory fit to the data has been obtained by a model dielectric function containing the Extended Drude term with a non-constant density of states (see Methods) and six Lorentz oscillators (with central energies: 1.53 eV, 1.99 eV, 2.72 eV, 3.85 eV, 4.36 eV and 5.05 eV), accounting for interband and charge transfer optical transitions. In the Extended Drude Model, we consider the same Glue Function $\Pi(\Omega)$ as the one determined from the fitting in the normal state (*T*=300 K) data (see (2)). Assuming $\Delta_{pg}$=40 meV (as determined from STS measurements, see (3)), the best fit to $R_{eq}(\omega)$ provides a value of $\widetilde{N}(0, T = 100\text{ K})$=0.7±0.05, (see Supplementary Note 2) in quantitative agreement with the results reported in (4).

# Supplementary Note 2

## The Non-Constant Density of States and the Photoinduced gap filling

In Supplementary Fig. S2(a) we report (dark green line) the normalized density of states obtained from fitting the Extended Drude Model with a non-constant density of states (see Methods and Supplementary Note 1) to the reflectivity data at equilibrium, reported in Supplementary Fig. S1. The best fit provides $\widetilde{N}_{eq}(0, T=100\,\text{K})=0.7\pm0.05$, when the pseudogap width value is fixed to $\Delta_{pg}=40$ meV (3). After the impulsive pump excitation, a transient gap filling can occur. In Supplementary Fig. S2(b) we report the $\delta R/R(\omega)$ signal arising from a partial filling of the pseudogap, that is accounted for by a partial quench of the density of states, i.e., $\widetilde{N}_{neq}(0, T=100\,\text{K}) - \widetilde{N}_{eq}(0, T=100\,\text{K}) = 0.2 \pm 0.02$.

## Supplementary Note 3

**Isosbestic Points: the link between Electronic Scattering Rate and Optical Properties**

In general, the transient variation of the electronic properties after the interaction with the pump pulse, modifies the optical properties over a very broad frequency range. Here we show how broadband time-resolved optical spectroscopy can be used to measure the instantaneous optical scattering rate with 100 fs time resolution. This argument is based on the properties of the so-called '*isosbestic points*' (defined as the crossing frequency ($\omega$) of all the functions $f(\omega,g)$ with different values of the parameter $g$), discussed in detail in a recent paper (5). In the vicinity of the crossing point, $\partial f/\partial g \simeq 0$ and the function can be expanded with respect to the parameter considered, establishing a simple dependence of $f(\omega,g)$ on the parameter $g$. As an example, we consider the variation of the scattering rate $\gamma$ on the prototypical reflectivity $R(\omega)$ of a metal. Within the simple Drude Model, recalling that:

$$R(\omega) = \left| (\sqrt{1-\varepsilon(\omega)})/(\sqrt{1+\varepsilon(\omega)}) \right|^2$$

$$\varepsilon_D(\omega) = \varepsilon_{\inf} - \frac{\omega_p^2}{\omega^2 + i\gamma\omega}$$

the width of the plasma edge is proportional to the scattering rate $\gamma$, defined as the inverse of the scattering time $\tau$. This is shown in Supplementary Fig. S3a, where a frequency-independent value $\tau$=5 fs is considered. The values of the other parameters used in the simulation are: $\varepsilon_{\inf}$=3, $\omega_p$=1.1 eV, $\gamma$=0.12 eV$\approx$(5 fs)$^{-1}$. A 20% increase (decrease) of the scattering rate $\gamma$ results in a broadening (narrowing) of the plasma edge. An isosbestic point (see Supplementary Fig. S3a) is found at the frequency $\widetilde{\omega}$~0.7 eV. Close to $\widetilde{\omega}$ the reflectivity variation can be expanded as $\delta R(\omega, \gamma)=\partial R/\partial\gamma(\omega)\delta\gamma$, where $\partial R/\partial\gamma$>(<)0 for $\omega$>(<) $\widetilde{\omega}$. Therefore, the effect of a pump-induced modification of the scattering rate can be readily inferred by measuring $\delta R/R(\omega,t)$ over a broad frequency range across $\widetilde{\omega}$, as shown in Supplementary Fig S3b.

# Supplementary Note 4

## Relaxation Dynamics in the Pseudogap

The relaxation time of the pump-probe measurements has been determined by fitting a single exponential decay to the pump-probe trace measured at $T$=100 K on the sUD-YBi2212 sample, excited with a pump fluence equal to 10±2 $\mu$J/cm$^2$. The result is reported in Supplementary Fig. S4. A decay time $\tau$=550±50 fs ($\hbar\tau^{-1}=\gamma=1.82$ ps$^{-1}$) has been obtained. The same decay time has been measured increasing the pump fluence up to 100 $\mu$J/cm$^2$.

## Supplementary Note 5

**ps-timescale Reflectivity Variation**

The $\delta R/R(\omega, t=3\text{ ps})$ measured at $T=100$ K on the OP sample is reported in Supplementary Fig. S5 (red dots). In this case, the measured frequency-dependent reflectivity variation is the opposite than that measured at t=100 fs (see Fig. 3a in the main paper). $\delta R/R(\omega, t=3\text{ ps})$ can be easily reproduced (yellow line) by assuming $\tilde{N}_{neq}(0, T=100\text{ K}) - \tilde{N}_{eq}(0, T=100\text{ K}) \simeq 0$ and a relative increase of the scattering rate $\delta\gamma_\infty/\gamma_\infty \approx 10^{-4}$. The long-time increase of the scattering rate is compatible with a local effective heating of $\delta T=0.6$ K, that corresponds to the effective heating estimated considering the pump fluence (10 µJ/cm$^2$) of the experiment and the specific heat of the lattice.

## Supplementary Note 6

### $T^*_{neq}(p)$ from non-equilibrium spectroscopy

The onset temperature $T^*_{neq}$ of the pseudogap, as a function of the hole-doping concentration, has been determined from the analysis of the single-colour (1.55 eV) pump-probe measurements. The appearance in the $\delta R/R(t)$ signal of a negative component with the relaxation time $\tilde{t}$ =600 fs is the consequence of the decrease of the electronic scattering rate as the non-thermal antinodal population is created by the pump pulse. The $\delta R/R(t)$ data at different temperatures and hole-doping concentrations are reported in Supplementary Fig. S6. The function $f(t)$, defined as the sum of three exponentials convoluted with a Gaussian representing the experimental resolution, has been fitted to the data:

$$f(t) = I_1 \cdot e^{-t/\tau_1} + I_2 \cdot e^{-t/\tau_2} + I_3 \cdot e^{-t/\tau_3}$$

The two positive components are needed to reproduce the temporal dynamics in the normal state. They exhibit two distinct decay times: a fast component of ~100 fs ($I_1$), and a slower one, with a decay time of ~1 ps ($I_2$). These timescales are related to the coupling of the electrons with two different subsets of phonons, as previously shown (2, 6). The third negative component ($I_3$) accounts for the transient decrease of the scattering rate in the pseudogap. The temperatures at which a non-zero value of $I_3$ appears are: $T^*$=240, 165 and 110 K for samples with $p$=0.13, 0.16 and 0.18, respectively. A zero value of $I_3$ is obtained at all the temperatures larger than $T_c$ for dopings $p$>0.18.

## Supplementary Note 7

**The crossing point of $T^*_{neq}(p)$ and the superconducting dome**

In this section, we discuss the measurements performed with the time-resolved optical spectroscopy on two overdoped samples: sOD ($p$=0.18, $T_c$=94 K) and OD ($p$=0.2, $T_c$=83 K). In the sOD sample, the $\delta R/R(\omega,t=100$ fs) signal above 1.4 eV evolve from positive in the normal state to negative in the pseudogap state at about 100 K (see Supplementary Fig. S7). In the OD sample, the pseudogap-related negative $\delta R/R(\omega,t=100$ fs) signal above 1.4 eV, is never detected for all the temperature larger than $T_c$ (see Supplementary Fig. S8). This result demonstrates that the $T^*_{neq}(p)$ line meets the superconducting dome at a doping concentration of 0.18<$p$<0.2.

## Supplementary Note 8

**Measurements on Hg1201**

In this section, we discuss the results of time- and frequency-resolved reflectivity measurements on $HgBa_2CuO_{4+\delta}$ (Hg1201) samples. Supplementary Figures S9 and S10 display the $\delta R/R(\omega,t)$ maps measured on underdoped (UD-Hg1201, $p\approx0.06$ $T_c$=55 K) and slightly-underdoped (sUD-Hg2101, $p\approx0.12$ $T_c$=95 K) Hg1201 crystals. The measurements have been performed in the same experimental conditions as the ones described for Y-Bi2212 samples. The spectral traces taken at t=0 at different temperatures are shown in the panels on the right. The $\delta R/R(\omega,t=100$ fs) signal at $T$=150 K on the sUD-Hg2101 is very similar to the signal measured on underdoped YBi2212 samples. The negative reflectivity variation at photon energies larger than 1.3 eV is the fingerprint of the transient decrease of the scattering rate, related to the pseudogap state. The signal related to the pseudogap gradually disappears as the temperature is increased up to $T$=300 K. At $T$=300 K the $\delta R/R(\omega,t=100$ fs) is positive for all photon energies and can be reproduced by simply assuming an increase of the scattering rate, in agreement with the results on Y-Bi2212 at $T$=300 K. On UD-Hg1201 the $\delta R/R(\omega,t=100$ fs) signal at $T$=150 K is very similar and exhibit a negative variation above 1.3 eV at all the temperatures from $T_c$ to 350 K. Supplementary Figure S11 reports the high-resolution, single colour pump-probe measurements performed at different sample temperatures on both the UD and the sUD Hg1201 crystals. In UD-Hg1201 the negative signal characteristic of the pseudogap state persists till the highest temperature is reached (350 K). Further increasing of the temperature leads to a permanent damage of the crystal. This demonstrates that the UD-Hg1021 sample is well within the pseudogap state at $T$=300 K.

## Supplementary References

1. van der Marel, D., Molegraaf, H. J. A., Zaanen, J., Nussinov, Z., Carbone, F., Damascelli, A., Eisaki, H., Greven, M., Kes, P. H. & Li, M. Quantum critical behaviour in a high-$T_c$ superconductor. *Nature* **425**, 271 (2003).
2. Dal Conte, S., Giannetti, C., Coslovich, G., Cilento, F., Bossini, D., Abebaw, T., Banfi, F., Ferrini, G., Eisaki, H., Greven, M., Damascelli, A., van der Marel, D. & Parmigiani, F. Disentangling the Electronic and Phononic Glue in a High-$T_c$ Superconductor. *Science* **335**, 1600 (2012).
3. Renner, Ch., Revaz, B., Genoud, J.-Y., Kadowaki, K. & Fischer, O. Pseudogap Precursor of the Superconducting Gap in Under- and Overdoped $Bi_2Sr_2CaCu_2O_{8+\delta}$. *Phys. Rev. Lett.* **80**, 149 (1998).
4. Hwang, J., Electron-boson spectral density function of underdoped $Bi_2Sr_2CaCu_2O_{8+\delta}$ and $YBa_2Cu_3O_{6.5}$. *Phys. Rev. B* **83**, 014507 (2011).
5. Greger, M., Kollar, M. & Vollhardt, D. Isosbestic points: How a narrow crossing region of curves determines their leading parameter dependence. *Phys. Rev. B* **87**, 195140 (2013).
6. Perfetti, L., Loukakos, P. A., Lisowski, M., Bovensiepen, U., Eisaki, H. & Wolf, M. Ultrafast Electron Relaxation in Superconducting $Bi_2Sr_2CaCu_2O_{8+\delta}$ by Time-Resolved Photoelectron Spectroscopy. *Phys. Rev. Lett.* **99**, 197001 (2007).